\documentclass[12pt]{article}
\usepackage[makeroom]{cancel}
\usepackage{subfigure}
\usepackage{float}
\usepackage{xspace}
\usepackage{color}
\usepackage{cite}
\usepackage{amssymb} 
\usepackage{amsmath}
\usepackage{amsfonts}
\usepackage{mathtools}

\usepackage{bm}
\usepackage{graphicx}        

\usepackage[unicode=true,plainpages=false]{hyperref}
\hypersetup{colorlinks=false,pdfborder={0 0 0.1},linkcolor=magenta,anchorcolor=magenta,urlcolor=blue,citecolor=blue,pdftitle={},pdfauthor={Alexander Litvinenko}}

\DeclareMathAlphabet{\mathitbf}{OML}{cmm}{b}{it}
\def\jmath{j}
\newcommand{\mynu}{\nu}
\newcommand{\myalpha}{\varphi_x}
\newcommand{\mybeta}{\varphi_y}
\newcommand{\mygamma}{\varphi_z}

\newcommand{\CMLMC}{\mbox{CMLMC}\xspace}
\newcommand{\MLMC}{\mbox{MLMC}\xspace}

\newcommand{\E}[1]{{\ensuremath{\mathrm{E}}\mspace{-2mu}\left[#1\right]}}
\newcommand{\prob}[1]{{\ensuremath{\mathrm{P}}\mspace{-2mu}\left[#1\right]}}
\newcommand{\var}[1]{{\ensuremath{\mathrm{Var}}\mspace{-2mu}\left[#1\right]}}

\providecommand{\tol}{\mathrm{TOL}}

\providecommand{\work}{\ensuremath{W}}

\providecommand{\est}[1]{\overset{\sim}{#1}}
\providecommand{\Order}[1]{ {\ensuremath{ \mathcal O\left( #1 \right)}} }

\def\eps{\varepsilon}

\providecommand{\tol}{\mathrm{TOL}}

\def\U{\mathit{U}}
\def\co{\alpha}
\def\mytheta{\vartheta}
\def\myphi{\varphi}

\def\sf{\sin(\phi)}

\newcommand{\TOL}{\mbox{TOL}}

\makeindex

\newcommand{\EXP}[1]{\mathbb{E}\left(#1\right)}

\def\xib{\bm{\xi}}

\newtheorem{theorem}{Theorem}

\DeclareMathAlphabet{\Bi}{OT1}{cmm}{b}{it}

\begin{document}
\title{Computation of Electromagnetic Fields Scattered From Objects With Uncertain Shapes Using Multilevel Monte Carlo Method}
\author{Alexander~Litvinenko\footnote{Corresponding author}, Abdulkadir~C.~Yucel,~Hakan~Bagci,\\ ~Jesper~Oppelstrup,~Eric~Michielssen,~and~Ra\'ul~Tempone\footnote{ 
A.~Litvinenko, H. Bagci, and R. Tempone are with the Strategic Research Initiative - Uncertainty Quantification (SRI-UQ) Center, Division of Computer, Electrical, and Mathematical Science and Engineering (CEMSE), King Abdullah University of Science and Technology (KAUST), Thuwal, 23955, Saudi Arabia (e-mails: \{alexander.litvinenko, raul.tempone, hakan.bagci\}@kaust.edu.sa).
A.~C.~Yucel is with the School of Electrical and Electronics Engineering, Nanyang Technological University, Singapore 639798 (e-mail: acyucel@ntu.edu.sg).
J.~Oppelstrup is with the Department of Mathematics, KTH Royal Institute of Technology, Stockholm, Sweden (e-mail: jespero@kth.se).
E.~Michielssen is with the Department of Electrical Engineering and Computer Science, University of Michigan, Ann Arbor, MI 48109, US (e-mail: emichiel@umich.edu).}}
\maketitle
\begin{abstract}
Computational tools for characterizing electromagnetic scattering from objects with uncertain shapes are needed in various applications ranging from remote sensing at microwave frequencies to Raman spectroscopy at optical frequencies. Often, such computational tools use the Monte Carlo (MC) method to sample a parametric space describing geometric uncertainties. For each sample, which corresponds to a realization of the geometry, a deterministic electromagnetic solver computes the scattered fields. However, for an accurate statistical characterization the number of MC samples has to be large. In this work, to address this challenge, the continuation multilevel Monte Carlo (\CMLMC) method is used together with a surface integral equation solver.
The \CMLMC method optimally balances statistical errors due to sampling of 
the parametric space, and numerical errors due to the discretization of the geometry using a hierarchy of discretizations, from coarse to fine.
The number of realizations of finer discretizations can be kept low, with most samples 
computed on coarser discretizations to minimize computational cost.
Consequently, the total execution time is significantly reduced, in comparison to the standard MC scheme.
\end{abstract}



\section{Introduction}
\label{sec:Intro}

Numerical methods for predicting radar and scattering cross sections (RCS and SCS) of complex targets find engineering applications ranging from microwave remote sensing soil/ocean surface and vegetation~\cite{tsang2013, tsang2017} to enhancing Raman spectroscopy using metallic nanoparticles~\cite{wang2005,talley2005}. When the target size is comparable to or larger than the wavelength at the operation frequency, the scattered field is a strong function of the target shape. However, in many of the applications, whether the target is a (vegetated) rough surface or a nanoparticle, its exact shape may not be known and has to be parameterized using a stochastic/statistical model. Consequently, computational tools, which are capable of generating statistics of a quantity of interest (QoI) (RCS or SCS in this case) given a geometry described using random variables/parameters, are required.

These computational tools often use the Monte Carlo (MC) method~\cite{stoll1984 , vesperinas1987, soto1989, sarabandi1993, wagner1997, jandhyala1998, tsang1998}. The MC method is non-intrusive and straightforward to implement; therefore, its incorporation with an existing deterministic EM solver is rather trivial. However, the traditional MC method has an error convergence rate of $\mathcal{O}(N^{-1/2})$~\cite{morokoff1995}, where $N$ is the number of samples in the parametric space used for describing the geometry. Provided more regularity of the QoI w.r.t. the geometry parameters, quasi-MC methods may have a better convergence rate, $\mathcal{O}(N^{-1})$ with a multiplicative $\log$-term that depends on the number of parameters~\cite{morokoff1995, QMLMC17}. Both the traditional MC and quasi-MC methods require large $N$ to yield accurate statistics of the QoI and are computationally expensive since the function evaluation at each sampling point, which corresponds to the computation of the QoI for one realization of the deterministic problem, requires the execution of a simulation. For practical real-life scattering scenarios, each of these simulations may take a few hours, if not a few days. 

In recent years, schemes that make use of surrogate models have received significant attention as potential alternatives to the MC method~\cite{xiu2007, xiu2005, foo2008, foo2010, ma2010, Giraldi2014, LitvSampling13}. The surrogate model is generated using the values of the QoI that are computed by the simulator at a small number of ``carefully selected'' (collocation) points in the parametric space. The surrogate model is then probed using the MC method to obtain statistics of the QoI. Consequently, the surrogate model-based schemes are more efficient than the MC method that operates directly on the computationally expensive simulator. The efficiency and accuracy of these schemes can be improved using a refinement strategy that adaptively divides the parametric space into smaller domains, in each of which a separate collocation scheme is used~\cite{foo2008, foo2010}. Additionally, if the QoI can be approximated in terms of low-order contributions from only a few of the parameters, one can use high dimensional model representation expansions to accelerate the generation of the surrogate model~\cite{ma2010}. The surrogate model-based schemes and  and their accelerated variants have been successfully applied to certain stochastic EM problems~\cite{xing2016, weng2015, austin2013, bagci2009, yucel2013a, yucel2013b, ochoa2013, yucel2015, gomez2015}. On the other hand, for the problem of scattering from geometries with uncertain shapes, all parameters and their high-order combinations contribute significantly to the QoI limiting the efficient use of a surrogate model~\cite{zeng2007, chauviere2006, chauviere2007}. 

Recently, the multilevel MC (\MLMC) methods have seen increasing use due to their efficiency, robustness, and simplicity~\cite{giles2008, giles2015, Barth2011, charrier2013, Cliffe2011, teckentrup2013, CMLMC}. The MLMC methods operate on a hierarchy of meshes and perform most of the simulations using low-fidelity models (coarser meshes) and only a few simulations using high-fidelity models (finer meshes). By doing so, their cost becomes significantly smaller than the cost of the traditional MC methods using only high-fidelity models. The continuation \MLMC (\CMLMC) algorithm~\cite{CMLMC} uses the \MLMC method to calibrate the average cost per sampling point and the corresponding variance using Bayesian estimation, taking particular notice of the deepest levels of the mesh hierarchy, to minimize the computational cost. To balance discretization and statistical errors, the \CMLMC method estimates how the discretization error and the computational cost depend on the mesh level and uses this information to select optimal numbers of levels and samples on each level.

In this work, a computational framework, which makes use of the \CMLMC method to efficiently and accurately characterize EM wave scattering from dielectric objects with uncertain shapes, is proposed. The deterministic simulations required by the \CMLMC method to compute the samples at different levels are carried out using the Poggio-Miller-Chan-Harrington-Wu-Tsai surface integral equation (PMCHWT-SIE) solver~\cite{mitchang1994}. The PMCHWT-SIE is discretized using the method of moments (MoM) and the iterative solution of the resulting matrix system is accelerated using a (parallelized) fast multipole method (FMM) - fast Fourier transform (FFT) scheme (FMM-FFT)~\cite{rokhlin1992, rokhlin1993, volakis2007, taboada2009, taboada2013, yucel2017a, yucel2017b}. Numerical results, which demonstrate the accuracy, efficiency, and convergence of the proposed computational framework, are presented. The developed computational framework is proven effective not only in studying the effects of uncertainty in the geometry on scattered fields but also in increasing the robustness of the FMM-FFT accelerated PMCHWT-SIE solver by testing its convergence for a large set of scenarios with deformed geometries and varying mesh densities, quadrature rules, iterative solver tolerances, and FMM parameters.  

\section{Formulation}
\label{sec:formulation}
\begin{figure}[t]
\center
\includegraphics[width=0.95\textwidth]{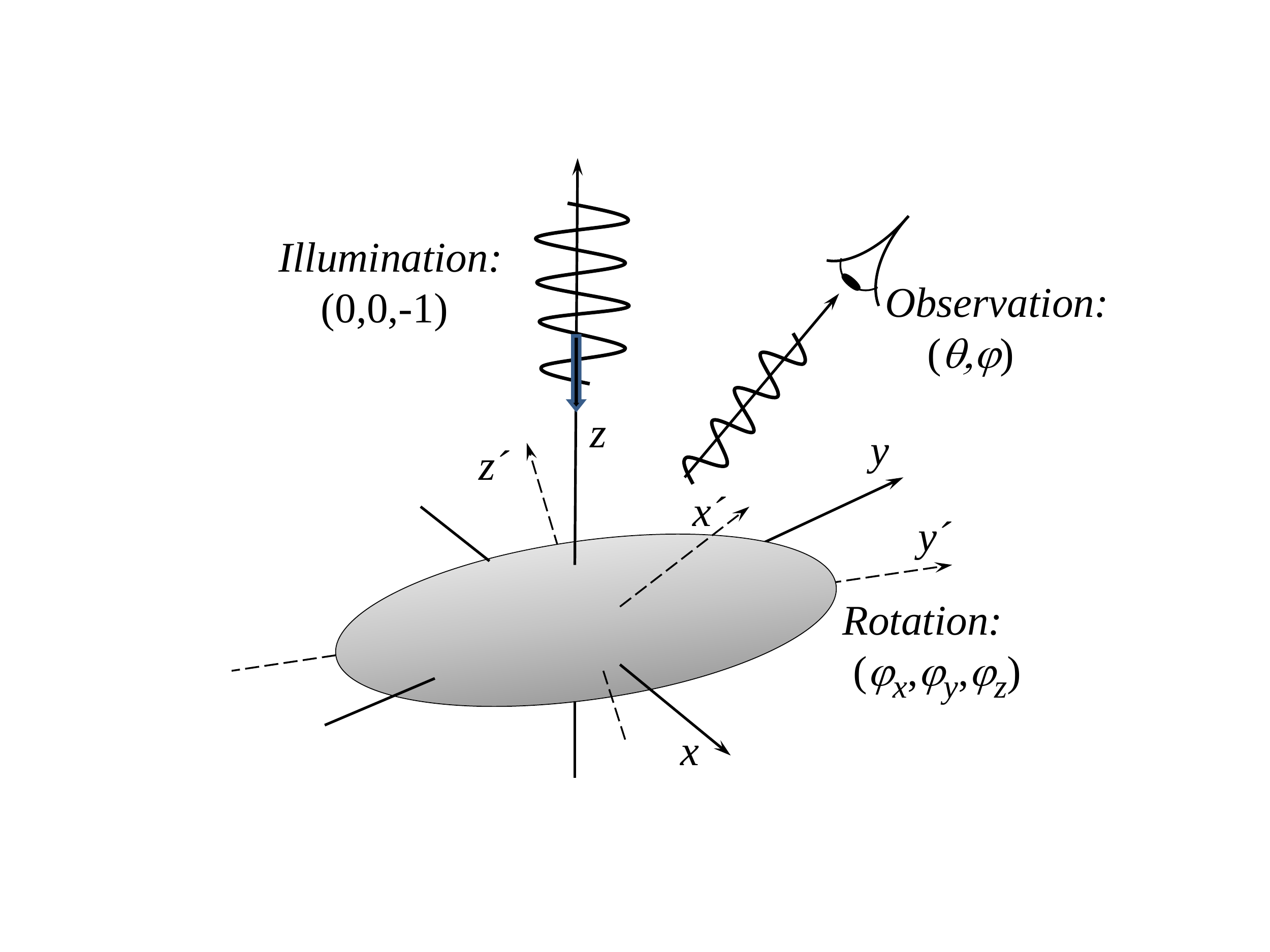}
\caption{Description of the scattering problem.}
\label{fig:scatter}
\end{figure}
This section describes a computational framework for characterizing scattering from dielectric objects with uncertain shapes (\ref{fig:scatter}). It is assumed that the shape of the scatterer can be parameterized by means of random variables. The QoI is the SCS over a user-defined solid angle (i.e., a measure of far-field scattered power in a cone). Section~\ref{sec:MLMC} describes the \MLMC and \CMLMC methods, Section~\ref{sec:RandomPerturb} formulates a scheme to parameterize and generate objects with uncertain shapes, and Section~\ref{sec:EC} outlines the FMM-FFT accelerated PMCHWT-SIE solver used for computing the scattering cross section of a given object.
\subsection{\MLMC and \CMLMC Methods}
\label{sec:MLMC}
This section describes elements of the \MLMC and \CMLMC methods relevant to the characterization of scattering from objects with uncertain shapes. A more in-depth description of these techniques can be found in~\cite{CMLMC}.

Let $\xib$ and $g(\xib)$ represent the vector of random parameters/variables and the QoI, respectively. The goal of the \MLMC method is to approximate the expected value, $\E{g}$, to a guaranteed tolerance $\TOL$ with predefined probability, and minimal computational cost. To achieve this, the method constructs a telescoping sum, defined over a sequence of mesh levels $\ell=0, \ldots, L$ as described next. 

Let $\{{P_\ell}\}_{\ell=0}^L$ and $\{h_{\ell}\}_{\ell=0}^L$ be sequences of meshes that discretize the randomly generated object's surface and their average element edge sizes, respectively. It is assumed $\{{P_\ell}\}_{\ell=0}^L$ are generated hierarchically with ${h_\ell=h_0 \beta^{-\ell}}$ for $h_0>0$ and a constant  $\beta > 1$. A method to such meshes for the purpose of illustrating the proposed technique is described in Section~\ref{sec:RandomPerturb}.

Let $g_\ell(\xib)$ represent the approximation to $g(\xib)$ computed using mesh $P_\ell$. The \MLMC method expresses $\E{g_L}$, the expected value of the most accurate approximation $g_L$, using 
\begin{equation}
\label{eq:EgL}
  \E{g_L} = \sum_{\ell=0}^L \E{G_{\ell}}
\end{equation}
where $G_\ell$ is defined as
\begin{align}
\label{eq:Gell}
    G_\ell &=
    \begin{cases}
        g_0 & \text{if } \ell=0 \\
        g_\ell- g_{\ell-1} & \text{if } \ell >0
    \end{cases}.
  \end{align}
Note that $g_\ell$ and $g_{\ell-1}$ are computed using the same input random parameter $\xib$.

In the telescoping sum~\eqref{eq:EgL}, the expected values in practice are replaced by sample averages, i.e., $\E{G_\ell} \approx \est G_\ell  =  M_\ell^{-1}\sum_{m=1}^{M_\ell} G_{\ell,m}$, where random variable $G_{\ell,m}$ have the same distribution as  
$G_{\ell}$ and are independent identically distributed (i.i.d.) samples. As $\ell$ increases, the variance
of $G_{\ell}$ decreases. As a result, the total computational cost can be reduced by approximating
$\E{G_{\ell}}$ with a smaller number of samples. 

The \CMLMC algorithm is an improved version of the MLMC method in that it approximates $\E{g}$ with a sequence of decreasing tolerances \cite{CMLMC}. In doing so, it continuously improves estimates of several problem-dependent parameters, while solving relatively inexpensive problems that by themselves would yield
large tolerances. The \CMLMC algorithm assumes that the convergence rates for the mean (weak convergence) and variance (strong convergence) follow
\begin{subequations}
\label{eq:q1q2}
\begin{align}
    \E{g-g_\ell} &\approx Q_W h_\ell^{q_1} \label{eq:weak_error_model} \\
    \var{g_\ell-g_{\ell-1}} &\approx Q_S h_{\ell-1}^{q_2} \label{eq:strong_error_model} 
\end{align}
\end{subequations}
for $Q_W\neq 0$, $Q_S>0$, $q_1>0$, and $0 < q_2 \leq 2 q_1$ \cite{CMLMC}.
For example, the \CMLMC algorithm estimates
$q_1 \approx 2$ and $q_2 \approx 4$, and 
$q_1 \approx 3$ and $q_2 \approx 5$ for the examples in Sections~\ref{ssec:Ex1} and~\ref{ssec:Ex2}, respectively. These parameter estimates are crucial to optimally distribute the
computational effort, as shown below. The total computational cost of the adaptive algorithm is close to that of the \MLMC method with correct values of parameters given a priori.

For the sake of completeness, the main ingredients of the $\CMLMC$ algorithm (described in full in~\cite{CMLMC}) are stated here. To estimate the number of samples, the algorithm invokes of \textit{`cost per sample'} and \textit{`total cost'} as described next.

The \CMLMC estimator for the QoI, $\mathcal{A}$, can be written as $ \mathcal{A} = \sum_{\ell=0}^L \est G_\ell$. Let the average cost of generating one sample of $G_{\ell}$ (cost of one deterministic simulation for one random realization) be

\begin{equation}
W_\ell \propto h_\ell^{-d\gamma} = h_0^{-d\gamma}\beta^{\ell d \gamma}
\label{eq:workpl}
\end{equation}
where $d$ is the spatial dimension and $\gamma$ is determined by the computational complexity of the deterministic solver. For the FMM-FFT-accelerated PMCHWT-SIE solver used here $d=2$ (only surfaces are discretized) and $\gamma\approx 1$ (Sections~\ref{sec:EC} and~\ref{ssec:Ex1}). Note that this solver calls for an iterative solution of the MoM system of equations (Section~\ref{sec:EC}), and that the cost of computing a sample of $G_\ell$ may fluctuate for different realizations depending on the number of iterations required. Finally, for the method used for generating $\{P_{\ell}\}_{l=0}^{L}$ (Section~\ref{sec:RandomPerturb}), $\beta=2$.

The total \CMLMC computational cost is
\begin{equation}
\label{eq:totalwork}
  \work = \sum_{\ell=0}^{L} M_\ell W_\ell.
\end{equation}
The estimator $\mathcal{A}$ satisfies a tolerance with a prescribed failure probability 
$0 < \mynu \leq 1$, i.e.,
\begin{align}
  \label{eq:goal}
  \prob{\left|\E{g}-\mathcal{A}\right| \leq \tol} &\geq 1-\mynu
\end{align}
while minimizing $\work$. The total error is split into bias and statistical error,
\begin{align*}
    \left|\E{g}-\mathcal{A}\right| \leq \underbrace{\left|\E{g-\mathcal{A}}\right|}_{\text{Bias}}
    + \underbrace{\left|\E{\mathcal{A}}-\mathcal{A}\right|}_{\text{Statistical error}}
\end{align*}
where $\theta \in (0,1)$ is a splitting parameter, so that
\begin{equation}
 \tol =  \underbrace{(1-\theta) \tol}_{\text{Bias tolerance}} +
\underbrace{\hskip 1cm \theta \tol \hskip1cm }_{\text{Statistical error tolerance}}.
\end{equation}
The \CMLMC algorithm bounds the bias, $B = \left|\E{g-\mathcal{A}}\right|$, and the 
statistical error as
\begin{align}
    B =  \left|\E{g-\mathcal{A}}\right| &\leq (1-\theta)\tol  \\
  \left|\E{\mathcal{A}}-\mathcal{A}\right| &\leq \theta \tol
  \label{eq:stat_error}
\end{align}
where the latter bound holds with probability $1-\mynu$. Note that $\theta$ itself can be a variable~\cite{CMLMC}. 

To satisfy condition in \eqref{eq:stat_error} we require:
\begin{align}
  \label{eq:var_bound}
  \var{\mathcal{A}} & \leq \left(\frac{\theta \tol}{C_\mynu} \right)^2
\end{align}
for some given confidence parameter, $C_\mynu$, such that 
${\Phi(C_\mynu) = 1-\frac{\mynu}{2}}$, (see more in \cite{ErikOptGeom15}); 
here, $\Phi$ is the cumulative distribution function of a standard normal random variable.

By construction of the \MLMC estimator, $\E{\mathcal{A}}=\E{g_L}$, and by independence,
  $\var{\mathcal{A}} = \sum_{\ell=0}^L V_\ell M_\ell^{-1}$, where $V_\ell = \var{G_\ell}$.
Given $L$, $\TOL$, and $0 < \theta < 1$, and by minimizing $W$ subject to the statistical constraint 
\eqref{eq:var_bound} for
        $\left\{M_\ell\right\}_{\ell=0}^L$ gives the following optimal 
        number of samples per level $\ell$ (apply ceiling function to $ M_\ell$ if necessary):
 \begin{equation}
    M_\ell = 
    \left( \frac{C_{\mynu}}{\theta \tol} \right)^2 \sqrt{\frac{V_\ell}{W_\ell}} 
    \left( \sum_{\ell=0}^L \sqrt{V_\ell W_\ell} \right).
    \label{eq:optimal_ml}
\end{equation}
Summing the optimal numbers of samples over all levels yields the following expression for the total optimal computational cost in terms of $\tol$:
\begin{equation}
    \work(\tol, L) = \left( \frac{C_\mynu}{\theta \tol} \right)^2 
    \left(\sum_{\ell=0}^L \sqrt{V_\ell W_\ell}\right)^2.
    \label{eq:work_per_L}
\end{equation}
%
%
%
The total cost of the \CMLMC algorithm can be estimated using Theorem~\ref{thm:costMLMC} below \cite{hoel2014implementation, hoel2012adaptive, charrier2013, Cliffe2011, giles2008}.
\begin{theorem}
\label{thm:costMLMC}
Let $d=\{1,2,3\}$ denote the problem dimension.
Suppose there exist positive constants $q_1,q_2,\gamma > 0$ such that
$q_1 \geq \frac{1}{2} \min(q_2, \gamma d)$, and
$\vert \E{g_{\ell}-g}\vert = \mathcal{O}(h_{\ell}^{q_1} )$, $\var{g_{\ell} - g_{\ell-1}} = \mathcal{O}(h^{q_2}_{\ell})$, and $W_{\ell} = \mathcal{O}(h^{-d\gamma}_{\ell})$.
Then for any accuracy $\TOL$ and confidence level $\nu$, $0< \nu \leq 1$,
there exist a deepest level
$L(\TOL)$ and number of realizations $\{M_{\ell}\left( \TOL \right )\}$ such that 
\begin{equation}
\lim_{\footnotesize{\TOL \to 0}}\inf \prob{\left( \vert \EXP{g} - \mathcal{A}\vert \leq \TOL \right)}\geq 1-\nu    
\end{equation}
with cost
\begin{align}
\label{eq:mlmc_iso_work} 
W(\TOL)\leq
      \begin{cases}
        \Order{\tol^{-2}}, & q_2 > d\gamma \\
        \Order{\tol^{-2} \left(\log(\tol^{-1})\right)^2}, & q_2= d\gamma \\
        \Order{\tol^{-\left(2 +\frac{d\gamma-q_2}{q_1}\right)}},  & q_2 < d\gamma \\
      \end{cases}.
    \end{align}
    \end{theorem}

This theorem shows that even in the worst case scenario, the \CMLMC algorithm has a lower computational cost than that of
the traditional (single level) MC method, which scales as $\mathcal{O}(\tol^{-2-d\gamma/q_1})$. 
Furthermore, in the best case scenario presented above, the computational 
cost of the \CMLMC algorithm scales as $\Order{\tol^{-2}}$, i.e. identical to that of the MC method assuming that the cost per sample is $\Order{1}.$ In other words, for this case, the \CMLMC algorithm can in effect remove the computational cost required by the discretization, namely  $\mathcal{O}(\tol^{-d\gamma/q_1})$.
\subsection{Random Shape Generation}
\label{sec:RandomPerturb}

This section describes a method to generate the sequence of meshes $\{{P_\ell}\}_{\ell=0}^L$ that are used in the numerical experiments (Section~\ref{sec:num_res}) to demonstrate the properties of the \CMLMC scheme. Alternative ways to generating random perturbations can be found in~\cite{litvinenko2013numerical, liu2017quantification, SchwabLang, Hcovariance,litvinenko2009sparse}. The method used here relies on perturbing an initial mesh generated on a sphere and refining it  as level $\ell$ is increased as described next. 

First, the unit sphere is discretized using a mesh $P_0$ with triangular elements, then the perturbations are generated by moving the nodes of this mesh using 
\begin{align}
\label{eq:KLE}
v(\mytheta_m,\myphi_m)&\approx \tilde{v}(\mytheta_m,\myphi_m)+ \sum_{k=1}^{K} a_k\kappa_k(\mytheta_m,\myphi_m).
\end{align}
where $\mytheta_m$ and $\myphi_m$ are angular coordinates of node $m$, $v(\mytheta_m,\myphi_m)$ is its (perturbed) radial coordinate, and $\tilde{v}(\mytheta_m,\myphi_m)=1$~m is its (unperturbed) radial coordinate on the unit sphere (all in spherical coordinate system). Here, $\kappa_k(\mytheta,\myphi)$ are obtained from spherical harmonics by re-scaling their arguments and $a_k$, which satisfy $\sum_{k=1}^K a_k <0.5$, are uncorrelated random variables. 

For the numerical experiments considered in this work, $K=2$, and $\kappa_1(\mytheta,\myphi)=\cos(\co_1 \mytheta)$ and $\kappa_2(\mytheta,\myphi)=\sin(\co_2 \mytheta) \sin(\co_3\myphi)$, where $\co_1$, $\co_2$, and $\co_3$ are positive constants. If a $\kappa_k(\mytheta,\myphi)$ depends on $\myphi$, then its dependence on $\mytheta$ must vanish at the poles $\mytheta = \{0,\pi\}$. Therefore, $\co_2$ must be an integer. Additionally, $\co_3$ must be an integer to generate a smooth perturbation.  

The random surface generated using~\eqref{eq:KLE} is also rotated and scaled as described next. Let $\bar R_x(\myalpha)$, $\bar R_y(\mybeta)$, and $\bar R_z(\mygamma)$ represent the matrices defined to perform rotations around axes $x$, $y$, and $z$ by angles $\myalpha$, $\mybeta$, and $\mygamma$, respectively: 
\begin{align}
\nonumber \bar R_x(\myalpha)=\left[ \begin{array}{ccc}
1 & 0 & 0 \\
0 & \cos \myalpha & -\sin \myalpha \\
0 & \sin \myalpha & \cos \myalpha \end{array} \right]\\
\nonumber \bar R_y(\mybeta)=\left[ \begin{array}{ccc}
\cos \mybeta & 0 & \sin \mybeta \\
0 & 1 & 0 \\
-\sin \mybeta & 0 & \cos \mybeta \end{array} \right]\\
\nonumber \bar R_z(\mygamma)=\left[ \begin{array}{ccc}
\cos \mygamma & -\sin \mygamma & 0 \\
\sin \mygamma & \cos \mygamma & 0 \\
0 & 0 & 1 \end{array} \right].
\end{align}
Similarly, let $\bar L(l_x, l_y, l_z)$ represent the matrix defined to implement scaling along axes $x$, $y$, and $z$ by $l_x$, $l_y$, and $l_z$, respectively: 
\begin{align}
\nonumber \bar L(l_x, l_y, l_z)=\left[ \begin{array}{ccc}
1/l_x & 0 & 0 \\
0 & 1/l_y & 0 \\
0 & 0 & 1/l_z \end{array} \right].
\end{align}
Then, the coordinates for the ``rotated'' and ``scaled'' mesh $P_0$ are obtained after the application of~\eqref{eq:KLE}, rotation and scaling matrices:
\begin{align}
\left[\! \begin{array}{c}
x'_m \\ y'_m  \\ z'_m 
\end{array}\! \right]
\!=\!\bar L(l_x, l_y, l_z)\bar R_x(\myalpha) \bar R_y(\mybeta) \bar R_z(\mygamma)
\left[\! \begin{array}{c}
x_m \\ y_m  \\ z_m 
\end{array}\! \right].
\label{eq:rot_sca_s}
\end{align}
Here, $(x_m, y_m, z_m)$, and $(x'_m, y'_m, z'_m)$ are the coordinates of node $m$ before and after the transformation.

The random variables used in generating the final version of $P_0$ are the perturbation weights $a_k$, $k=1,\ldots, K$, the rotation angles $\myalpha$, $\mybeta$, and $\mygamma$, and the scaling factors $l_x$, $l_y$, and $l_z$, making the dimension of the stochastic space $K + 6$, i.e., random parameter vector 
\begin{equation*}
\xib=\{a_1,\ldots, a_K, \myalpha, \mybeta, \mygamma, l_x, l_y, l_z\}.
\end{equation*}

\begin{figure}[t!]
\center
\subfigure[]{
\includegraphics[width=0.45\textwidth]{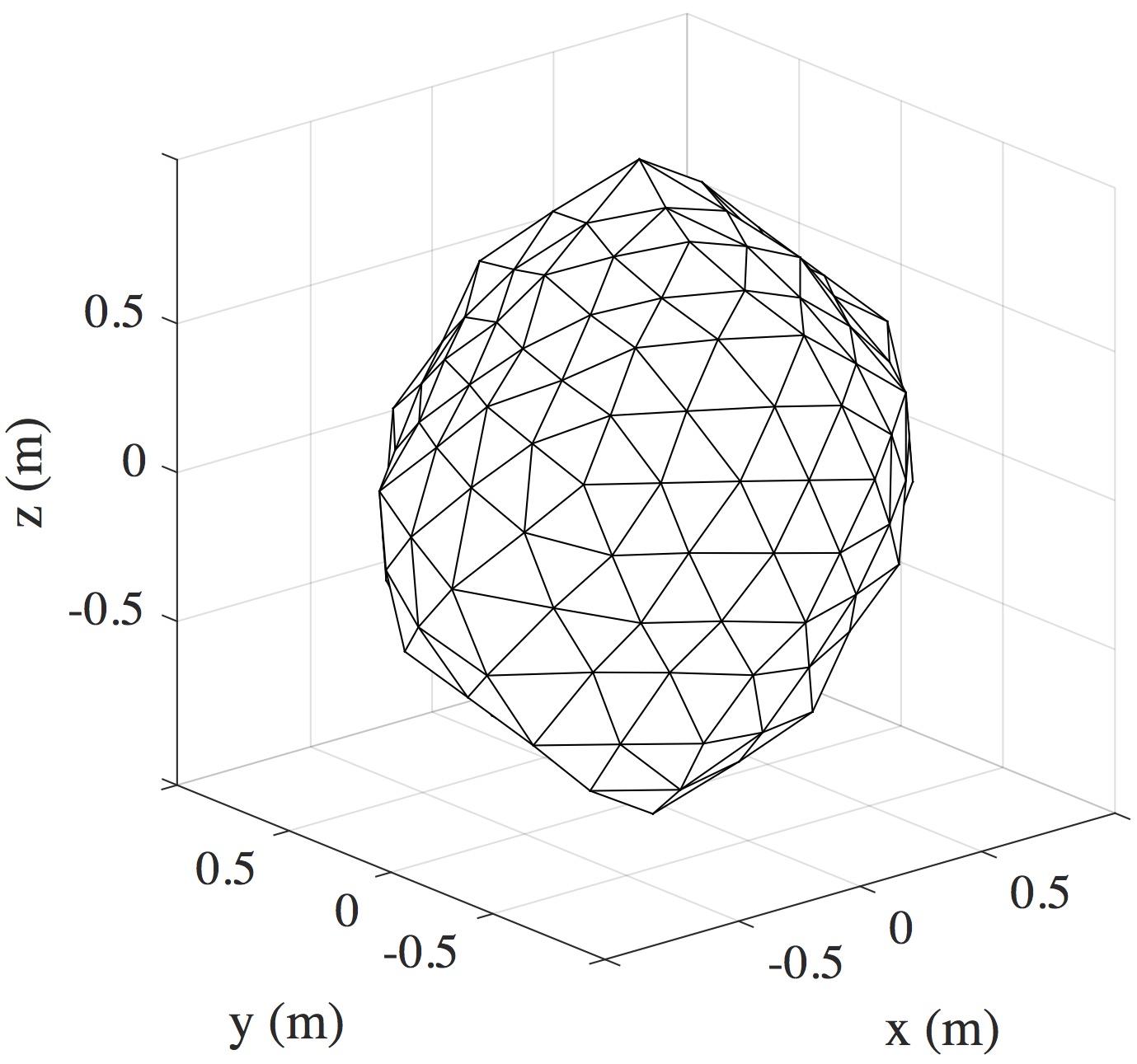}
\label{fig:mesh_A}}
\subfigure[]{
\includegraphics[width=0.45\textwidth]{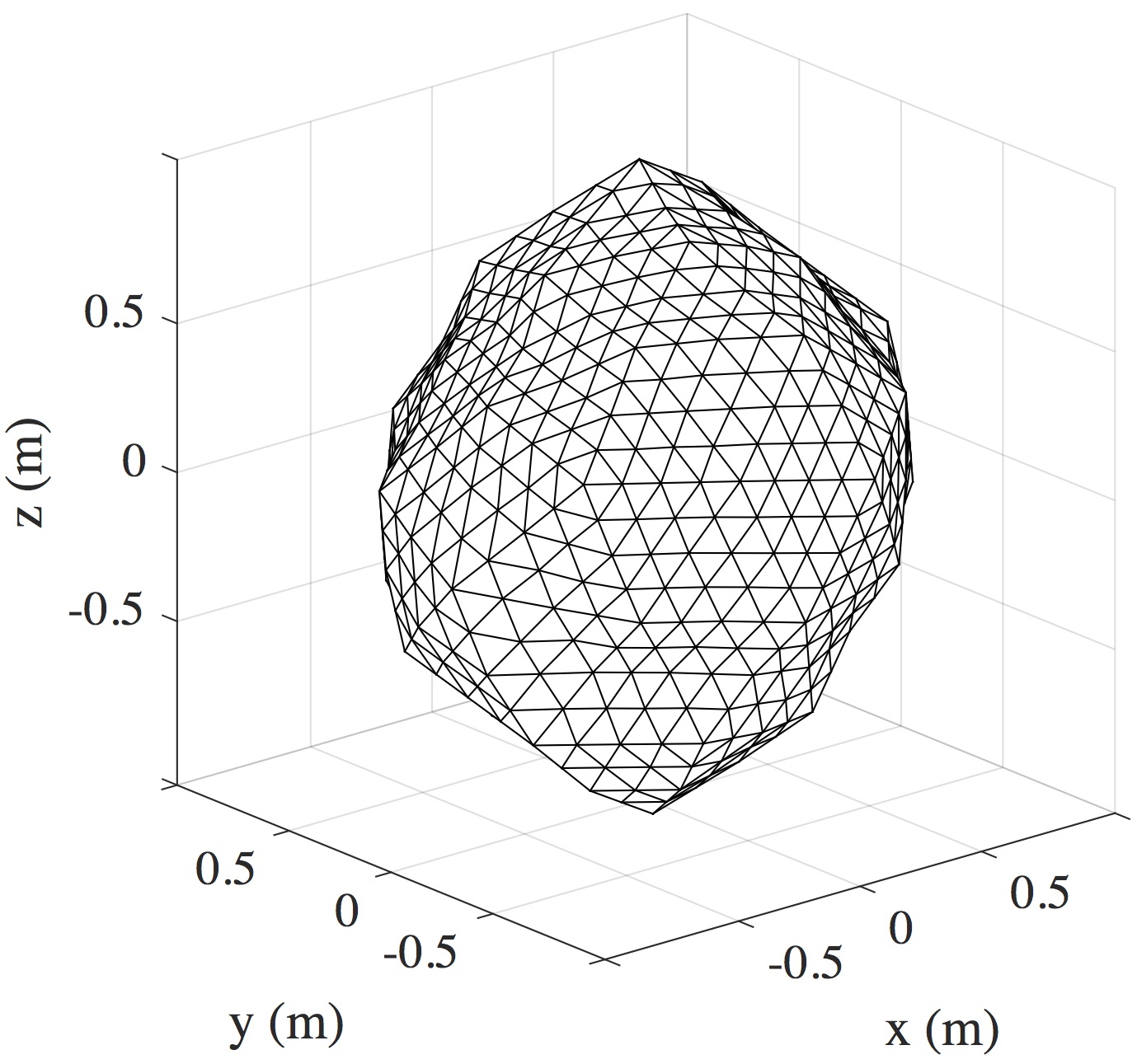}
\label{fig:mesh_B}}\\
\subfigure[]{
\includegraphics[width=0.45\textwidth]{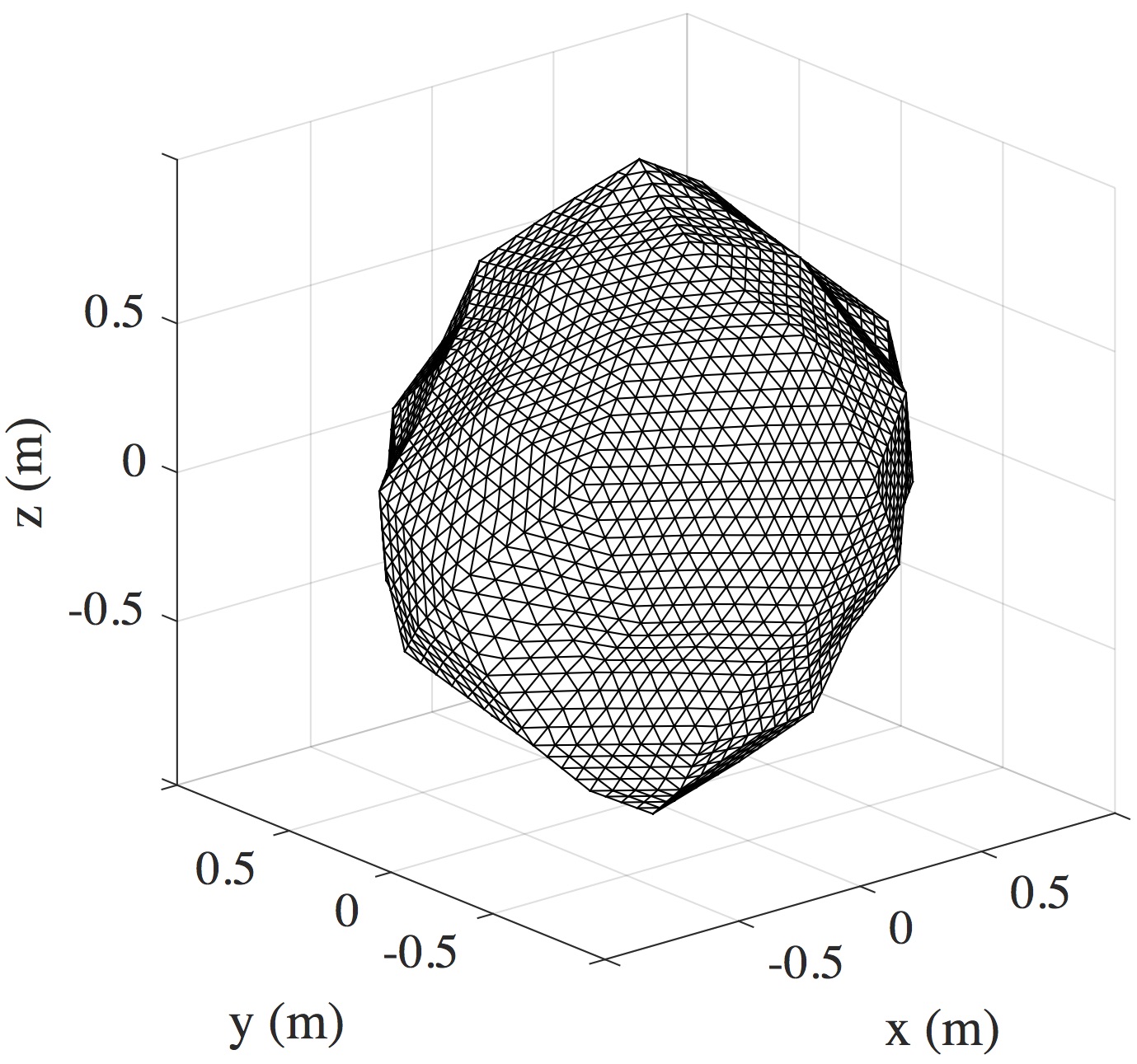}
\label{fig:mesh_C}}
\subfigure[]{
\includegraphics[width=0.45\textwidth]{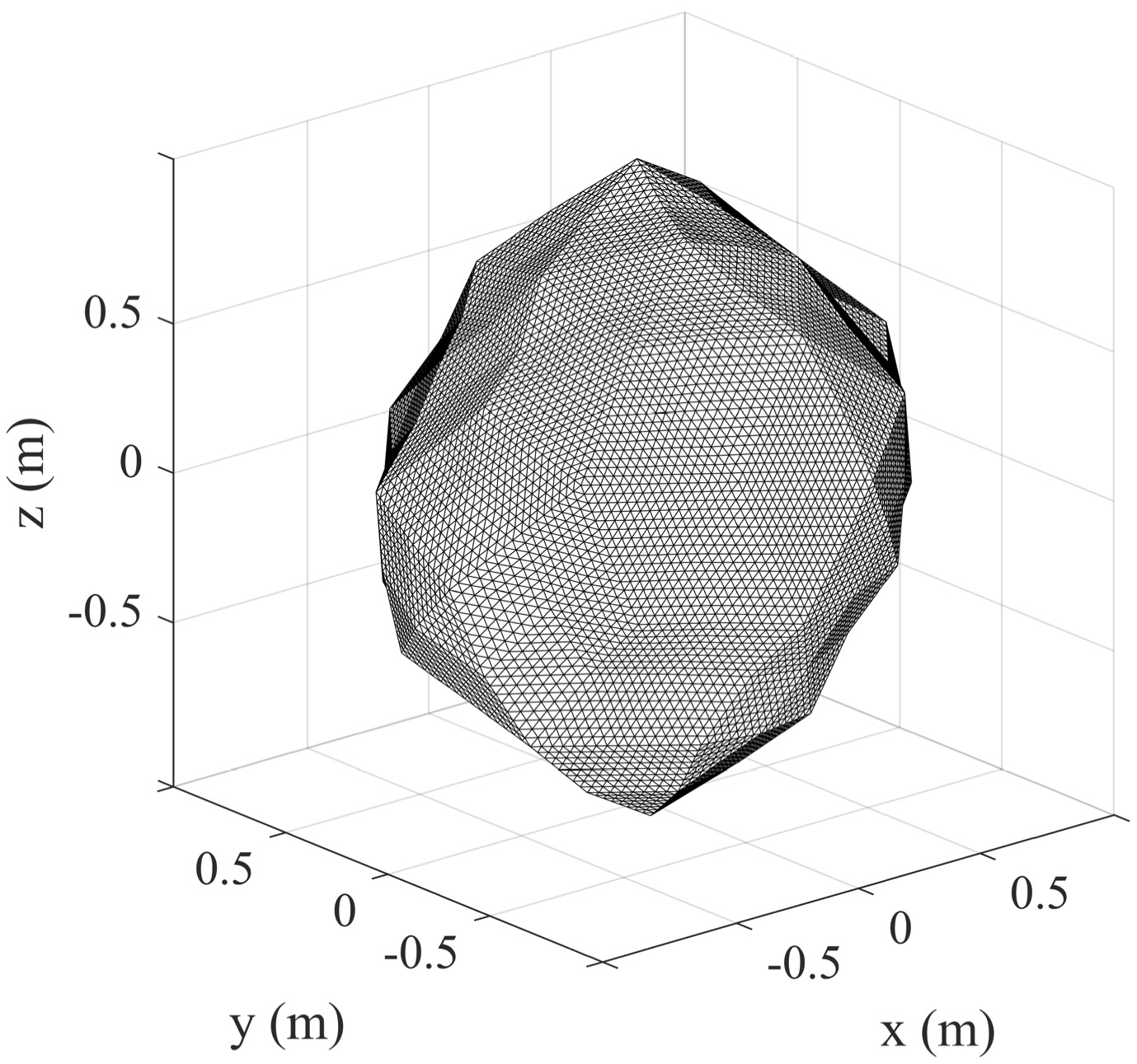}
\label{fig:mesh_D}}
\caption{An example of four nested meshes with (a) $320$, (b) $1280$, (c) $5120$, and (d) $20480$ triangular elements, which are generated for a perturbed shape with $\co_1=2$, $\co_2=3$, $\co_3=2$, $a_1=0.04$~m, $a_2=0.048$~m, $\myalpha=0.32$~rad, $\mybeta=0.88$~rad, $\mygamma = 0.81$ rad, $l_x=1.06$, $l_y=1.08$, and $l_z=1.07$.}
\label{fig:mesh}
\end{figure}

Note that $P_0$ is the coarsest mesh used in \CMLMC ($\ell=0$) (see Fig.~\ref{fig:mesh_A} for an example).
The mesh of the next level ($\ell=1$), $P_1$, is generated by refining each triangle of the perturbed $P_0$ into four (by halving all three edges and connecting mid-points) [Fig.~\ref{fig:mesh_B}]. The mesh at level $\ell =2$, $P_2$, is generated in the same way from $P_1$ [Fig.~\ref{fig:mesh_C}], and so on. All meshes $P_{\ell}$ at all levels $\ell=1,\ldots, L$ are nested discretizations of $P_0$. This method of refinement results in $\beta=2$ in \eqref{eq:workpl}. Note that no uncertainties are added on meshes $P_{\ell}$, $\ell>0$; the uncertainty is introduced only at level $\ell=0$. It is assumed that $P_0$ is fine enough to accurately represent the variations of the highest harmonic in the expansion used in~\eqref{eq:KLE}.
\subsection{Electromagnetic Solver}
\label{sec:EC}
This section briefly describes the PMCHWT-SIE solver used for the RCS and SCS of a dielectric scatterer with surface $S$. It is assumed that sources and fields are time-harmonic, i.e., their time dependence varies as ${e^{j\omega t}}$, where $t$ and $\omega$ are the time and angular frequency, respectively. 

Let ${V_1}$ and ${V_0}$ represent the space internal and external to $S$, respectively.  The permittivity, permeability, and characteristic impedance in ${V_i}$ , $i \in \{ 0,1\} $ are ${\varepsilon _i}$, ${\mu _i}$, and ${\eta _i} = \sqrt {{\mu _i}/{\varepsilon _i}}$, respectively. It is assumed that the scatterer is excited by an external electromagnetic field with electric and magnetic components ${{\bf{E}}^{{\rm{inc}}}}({\bf{r}})$ and ${{\bf{H}}^{{\rm{inc}}}}({\bf{r}})$. Using the surface equivalence theorem and enforcing the tangential continuity of total electromagnetic fields on $S$ yield the PMCHWT SIE~\cite{mitchang1994}: 
\begin{align}
\label{eq:pmchwt1}{\bf{\hat n}}({\bf r}) \times &{\bf{E}}^{\rm{inc}}({\bf r})={\bf{\hat n}}({\bf r}) \times \Big\{\mathcal{L}_0[{\bf{J}}]({\bf r})+\mathcal{L}_1[{\bf{J}}]({\bf r})\Big.\\ \nonumber&-\mathcal{K}_0[{\bf{M}}]({\bf r})-\mathcal{K}_1[{\bf{M}}]({\bf r})\Big\}, {\bf r} \in S
\end{align}
\begin{align}
\label{eq:pmchwt2}{\bf{\hat n}}({\bf r}) \times &{\bf{H}}^{\rm{inc}}({\bf r}) ={\bf{\hat n}}({\bf r}) \times \Big\{\mathcal{K}_0[{\bf{J}}]({\bf r})+\mathcal{K}_1[{\bf{J}}]({\bf r})\Big.\\
\nonumber & + \eta_0^{-2}\mathcal{L}_0[{\bf{M}}]({\bf r})+\eta_1^{-2}\mathcal{L}_1[{\bf{M}}]({\bf r})\Big\}, {\bf r} \in S.
\end{align}
Here, ${\bf{\hat n}}({\bf{r}})$ is the outward pointing unit normal vector, and ${\bf{J}}({\bf{r}})$ and ${\bf{M}}({\bf{r}})$ represent equivalent electric and magnetic surface current densities on $S$. The integral operators $\mathcal{L}_i[.]({\bf r})$ and $\mathcal{K}_i[.]({\bf r})$ in~\eqref{eq:pmchwt1} and~\eqref{eq:pmchwt2} are
\begin{align}
\nonumber \mathcal{L}_i[{\bf{X}}]({\bf r}) &= j\omega {\mu _i}\int_S {\left[ {{\bf{\bar I}} + \frac{{\nabla \nabla '}}{{k_i^2}}} \right] \cdot {\bf{X}}({\bf{r'}}){g_i}({\bf{r}},{\bf{r'}})ds'}\\   
\nonumber \mathcal{K}_i[{\bf{X}}]({\bf r}) &= \nabla  \times \int_S {{\bf{X}}({\bf{r'}}){g_i}({\bf{r}},{\bf{r'}})ds'} 
\end{align}
where ${g_i}({\bf{r}},{\bf{r'}})=e^{- jk_i|{\bf{r}} - {\bf{r'}}|}/(4\pi|{\bf{r}} - {\bf{r'}}|)$ is the Green function of the Helmholtz equation in the unbounded medium with wave number ${k_i} = \omega \sqrt {{\varepsilon _i}{\mu _i}}$.

To numerically solve~\eqref{eq:pmchwt1} and~\eqref{eq:pmchwt2} for the unknowns ${\bf{J}}({\bf{r}})$ and ${\bf{M}}({\bf{r}})$, $S$ is discretized by triangular elements, and ${\bf{J}}({\bf{r}})$ and ${\bf{M}}({\bf{r}})$ are approximated as
\begin{align}
\label{eq:j_exp}{\bf{J}}({\bf{r}}) &= \sum\limits_{n = 1}^N {\bar I}_n{\bf{f}}_n({\bf{r}})\\
\label{eq:m_exp}{\bf{M}}({\bf{r}}) &= \sum\limits_{n = N + 1}^{2N} {\bar I}_n{\bf{f}}_{n-N}({\bf{r}})
\end{align}
where ${\bf{f}}_n({\bf{r}})$, $n = 1,...,N$, represent Rao-Wilton-Glisson basis functions~\cite{rao_rwg}, and $\overline{I}=[I_1,\ldots,I_{2N}]^{\rm{T}}$ is the vector of unknown coefficients. Substituting~\eqref{eq:j_exp} and~\eqref{eq:m_exp} into~\eqref{eq:pmchwt1} and~\eqref{eq:pmchwt2}, and testing the resulting equations with ${\bf{f}}_m({\bf{r}})$, $m = 1,...,N$, yield the MoM system of equations:
\begin{align}
\label{eq:mom} {\bar V} = {\bar Z}{\bar I}
\end{align} 
where the entries of the vector $\bar V$ and the MOM matrix $\bar Z$ are
\begin{align}
\label{eq:v_m} {\bar V}_m &= \left\{ {\begin{array}{*{20}{l}}
\!\!\!\!{\left\langle {{{\bf{f}}_m},{{\bf{E}}^{{\rm{inc}}}}} \right\rangle,} & {1\leq m \leq N}\vspace{0.2cm}\\ 
\!\!\!\!{\left\langle {{{\bf{f}}_{m-N}},{{\bf{H}}^{{\rm{inc}}}}} \right\rangle,} & {N+1\leq m \leq 2N}
\end{array}} \right.
\end{align}
\begin{align}
\label{eq:z_mn} \!\!\!{\bar Z}_{m,n} &= \left\{ {\begin{array}{*{20}{l}}
\!\!\!\!{\left\langle {{{\bf{f}}_m}, \mathcal {L}_0[{{\bf{f}}_n}] + \mathcal {L}_1[{{\bf{f}}_n}]} \right\rangle,}\\ \;\; {1\leq m \leq N}, {1\leq n \leq N}\vspace{0.2cm}\\
\!\!\!\!{ - \left\langle {{{\bf{f}}_{m-N}},\mathcal {K}_0[{{\bf{f}}_n}] +\mathcal {K}_1[{{\bf{f}}_n}]} \right\rangle,}\\ \;\; {N + 1 \leq m \leq 2N},{1\leq n \leq N}\vspace{0.2cm}\\
\!\!\!\!{\left\langle {{{\bf{f}}_m},\mathcal {K}_0[{{\bf{f}}_{n-N}}] + \mathcal {K}_1[{{\bf{f}}_{n-N}}]} \right\rangle,}\\ \;\; {1 \leq n \leq N},{N + 1 \leq n \leq 2N}\vspace{0.2cm}\\
\!\!\!\!{\left\langle {{{\bf{f}}_{m-N}},\eta_0^{-2}\mathcal {L}_0[{{\bf{f}}_n}] + \eta_1^{-2}\mathcal {L}_1[{{\bf{f}}_n}]} \right\rangle,}\\ \;\; {N+1 \leq m \leq 2N},{N + 1 \leq n \leq 2N}
\end{array}} \right..
\end{align}
Here, the inner product is $\left\langle {{{\bf{f}}_m},{\bf{a}}} \right\rangle  = \int_{{S}} {{{\bf{f}}_m}({\bf{r}}) \cdot {\bf{a}}({\bf{r}})d{\bf{r}}}$.

Matrix equation~\eqref{eq:mom} is solved iteratively for $\bar I$. The computational cost of multiplying $\bar Z$ with a trial solution vector scales as $\mathcal{O}({N^{{\rm{iter}}}}{N^2})$, where ${N^{{\rm{iter}}}}$ is the number of iterations required for the residual error to reach the desired level: typically ${N^{{\rm{iter}}}} \ll N$. Likewise, the storage costs of the unaccelerated/classical solution scale as $\mathcal{O}({N^2})$.

To minimize the computational and storage cost while executing the MLMC algorithm a fast multipole method (FMM)-fast Fourier transform (FFT) scheme is used. A detailed formulation of FMM and its extension FMM-FTT can be
found in~\cite{rokhlin1992, rokhlin1993, volakis2007, taboada2009, taboada2013, yucel2017a, yucel2017b}. The scheme encloses the scatterer in a fictitious box that is embedded into a uniform grid of smaller boxes. Two non-empty boxes (i.e., boxes containing at least a pair of patches) constitute a far-field pair if there is at least one box between them. Otherwise, they form a near-field pair. Interactions between basis and test functions in near-field pairs (and same boxes) are computed using~\eqref{eq:z_mn} and stored in matrix ${\bar Z}^{\rm{near}}$. The contribution of near- and self-interactions to the matrix-vector multiplication $\bar Z \bar I$ is calculated by simply multiplying ${\bar Z}^{\rm{near}}$ with $\bar I$. The interactions between basis and test functions in far-field pairs are represented using the radiation patterns of basis and test functions (sampled over a solid angle of a sphere)   and a translation operator and their contributions to $\bar Z \bar I$ are computed using the FMM-FFT scheme. The computational cost and memory requirement of the FMM-FFT accelerated iterative solution scale as $\mathcal{O}({N^{{\rm{iter}}}}{N^{4/3}}{\log ^{2/3}}N)$ and $\mathcal{O}({N^{4/3}}{\log ^{2/3}}N)$, respectively~\cite{yucel2017a, yucel2017b}. Note that these estimates are higher than those of the multilevel FMM~\cite{chew1995, chew1997}; the FMM-FFT scheme is preferred here since its parallel implementation is significantly simpler~\cite{volakis2007, taboada2009, taboada2013, yucel2017a, yucel2017b, ergul2008, ergul2009, fostier2008, fostier2013, liu2016a, liu2016b, liu2018}. The parallelization strategy implemented here uses a hybrid message passing interface/open multiprocessing (MPI/OpenMP) standard to uniformly distribute the memory and computational load among processors~\cite{volakis2007, taboada2009, taboada2013, yucel2017a, yucel2017b, ergul2008, ergul2009, fostier2008, fostier2013, liu2016a, liu2016b, liu2018}. 

Note that the computational cost estimate provided above is obtained under the assumption that the ratio of the wavelength to the (average) element edge length stays the same as $N$ is decreased/increased~\cite{volakis2007, taboada2009, taboada2013, yucel2017a, yucel2017b}. Within the \CMLMC algorithm, the mesh is refined from one level to the next while the frequency is kept constant. This means that the ratio of the wavelength to the edge length increases and the computational cost of the FFT-FMM-accelerated solver is expected to scale differently. Indeed, numerical experiments in Section~\ref{sec:num_res} show that the \CMLMC algorithm estimates the computational cost parameter as $\gamma \approx 1$ for objects comparable to the wavelength in size.

To compute the RCS and SCS, the scatterer is excited by a plane wave: ${\bf E}^{\rm inc}({\bf r}) ={\bf E}_0e^{-jk_0 {\bf \hat u}^{\rm inc}\cdot {\bf r}}$ and ${\bf H}^{\rm inc}({\bf r}) =({\bf \hat u}^{\rm inc} \times {\bf E}_0)/\eta_0e^{-j k_0 {\bf \hat u}^{\rm inc} \cdot {\bf r}}$, where ${\bf \hat u}^{\rm inc}$ is the direction of propagation. The unknown vector $\bar I$ is solved for under this excitation and the RCS ${\sigma^{\rm rcs}}(\mytheta ,\myphi )$ along direction ${\bf{\hat u}}(\mytheta ,\myphi ) = {\bf{\hat x}}\sin \mytheta \cos \myphi  + {\bf{\hat y}}\sin \mytheta \sin \myphi  + {\bf{\hat z}}\cos \mytheta $ is computed using~\cite{ishimaru1978}
\begin{equation}
\label{eq:sigma}
{\sigma^{\rm rcs}}(\mytheta ,\myphi ) = \frac{\big| {{\bf{F}}(\mytheta ,\myphi )} \big|^2}{4\pi  \big| {\bf E}_0 \big|^2 }.
\end{equation}
Here, ${{\bf{F}}(\mytheta ,\myphi )}$ is the scattered electric field pattern in the far field and computed using
\begin{align}
\nonumber
&{\bf{F}}(\mytheta ,\myphi ) =  - j\omega {\mu _0}{\bf{\hat u}} (\mytheta ,\myphi )\\
\nonumber &\times {\bf{\hat u}} (\mytheta ,\myphi )\times\sum\limits_{n = 1}^N {{{\bar I}_n}} \int_{{S_n}} {{{\bf{f}}_n}({\bf{r'}}){e^{j{k_0}{\bf{\hat u}}(\mytheta ,\myphi ) \cdot {\bf{r'}}}}d{\bf{r'}}}\\
\nonumber&- j{k_0}{\bf{\hat u}} (\mytheta ,\myphi )\times\sum\limits_{n = N + 1}^{2N} {{{\bar I}_n}} \int_{{S_{n-N}}}\!\!\! \!\!\!{{{\bf{f}}_{n-N}}({\bf{r'}}){e^{j{k_0}{\bf{\hat u}}(\mytheta ,\myphi ) \cdot {\bf{r'}}}}d{\bf{r'}}}.
\end{align}
The SCS ${C^{{\rm{sca}}}}(\Omega )$ is obtained by integrating $\sigma ^{\rm rcs}(\mytheta ,\myphi )$ over the solid angle $\Omega$~\cite{ishimaru1978}:
\begin{align}
\label{eq:scs}{C^{{\rm{sca}}}}(\Omega ) = \frac{1}{{4\pi }}\int_\Omega  {{\sigma^{\rm rcs}}(\mytheta ,\myphi )\sin \mytheta d\mytheta d\myphi}.	
\end{align}
The integral in~\eqref{eq:scs} is efficiently computed using the exact quadrature rule in~\cite{yucel_ms_2008}.   

\begin{figure}[htbp!]
\center
\subfigure[]{
\includegraphics[width=0.4\textwidth]{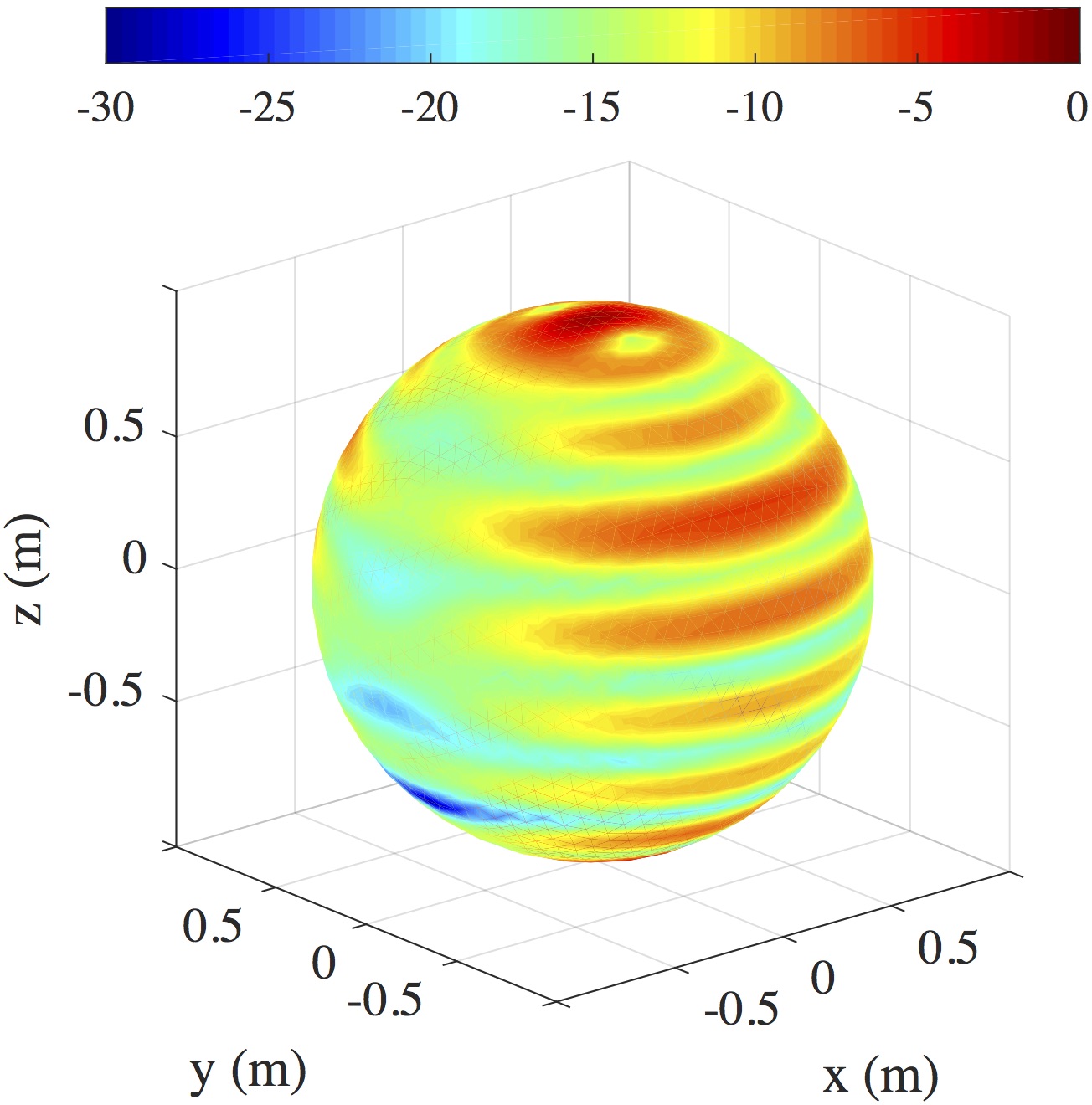}
\label{fig:elec_curr}}
\subfigure[]{
\includegraphics[width=0.4\textwidth]{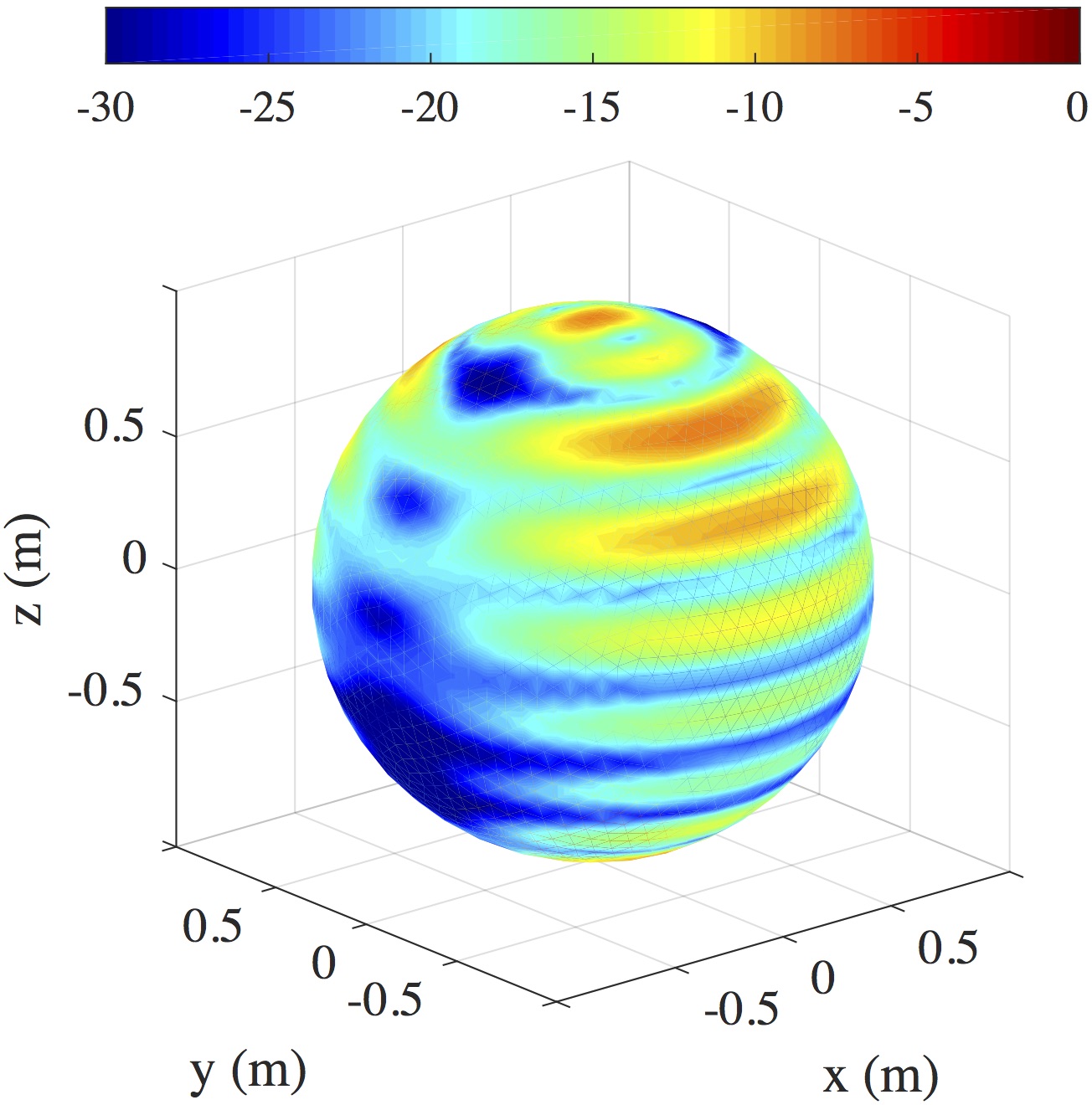}
\label{fig:mag_curr}}\\
\subfigure[]{
\includegraphics[width=0.4\textwidth]{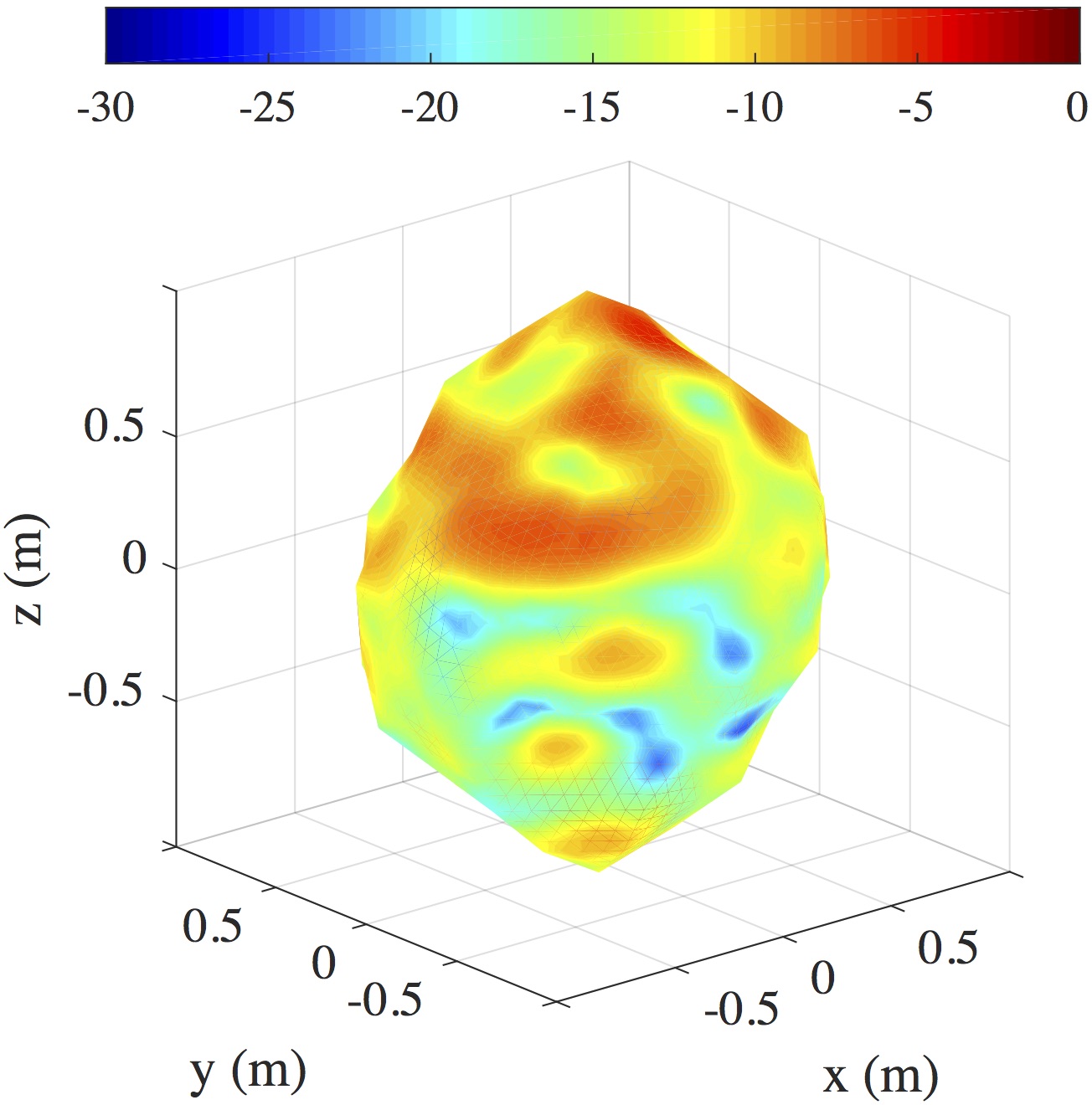}
\label{fig:elec_curr_pert}}
\subfigure[]{
\includegraphics[width=0.4\textwidth]{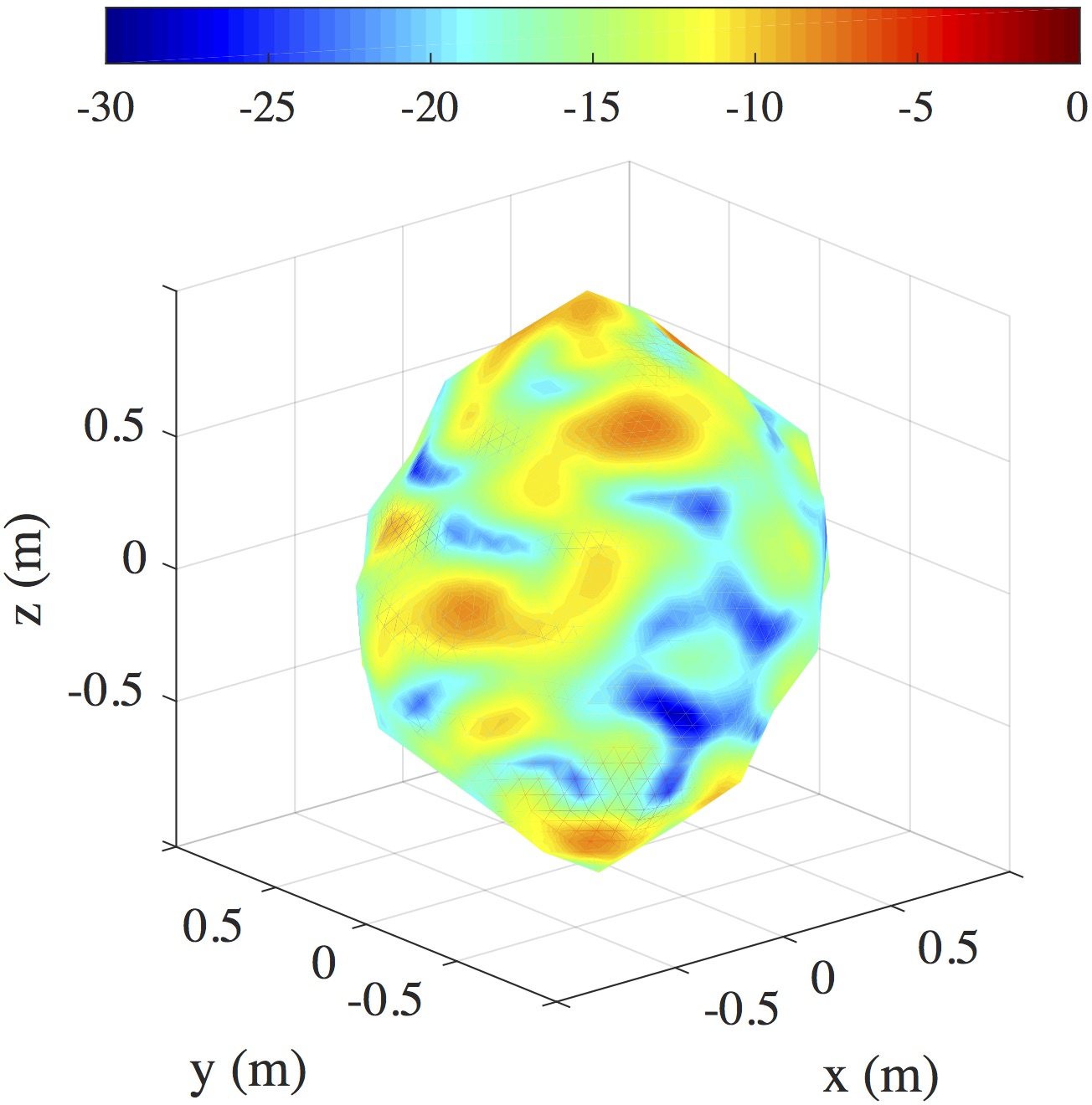}
\label{fig:mag_curr_pert}}
\caption{Amplitudes of (a) ${\bf J}({\bf r})$ and (b) ${\bf M}({\bf r})$ induced on the unit sphere under excitation by an $\hat {\bf x}$-polarized plane wave propagating in $-\hat {\bf z}$ direction at $300$~MHz. Amplitudes of (c) ${\bf J}({\bf r})$ and (d) ${\bf M}({\bf r})$ induced on the perturbed shape shown in Fig.~\ref{fig:mesh_C} under excitation by the same plane wave. For all figures, amplitudes are normalized to 1 and plotted in $dB$ scale.}
\label{fig:currents}
\end{figure} 
\begin{figure}[htbp!]
\center
\subfigure[]{
\includegraphics[width=0.45\textwidth]{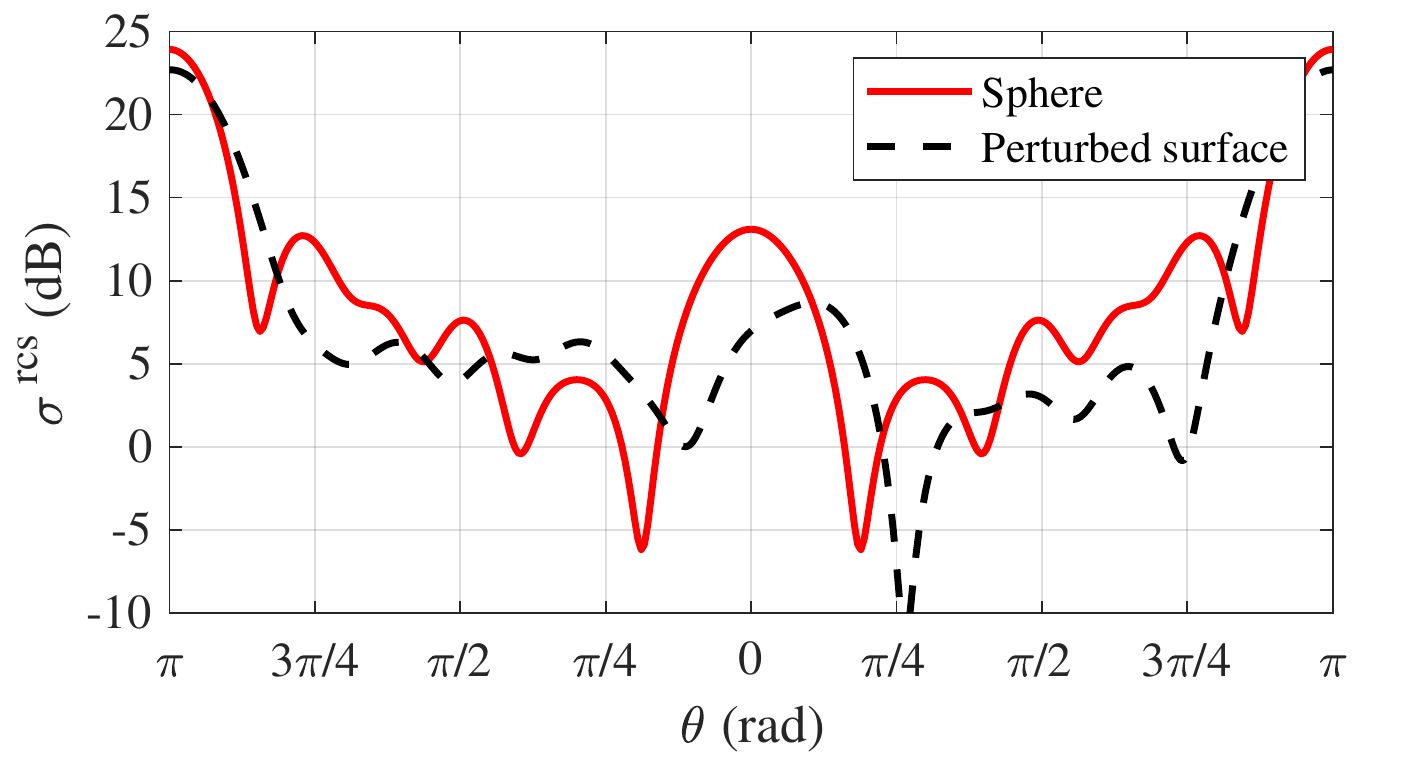}
\label{fig:rcs_xz}
}\vspace{-0.2cm}
\subfigure[]{
\includegraphics[width=0.45\textwidth]{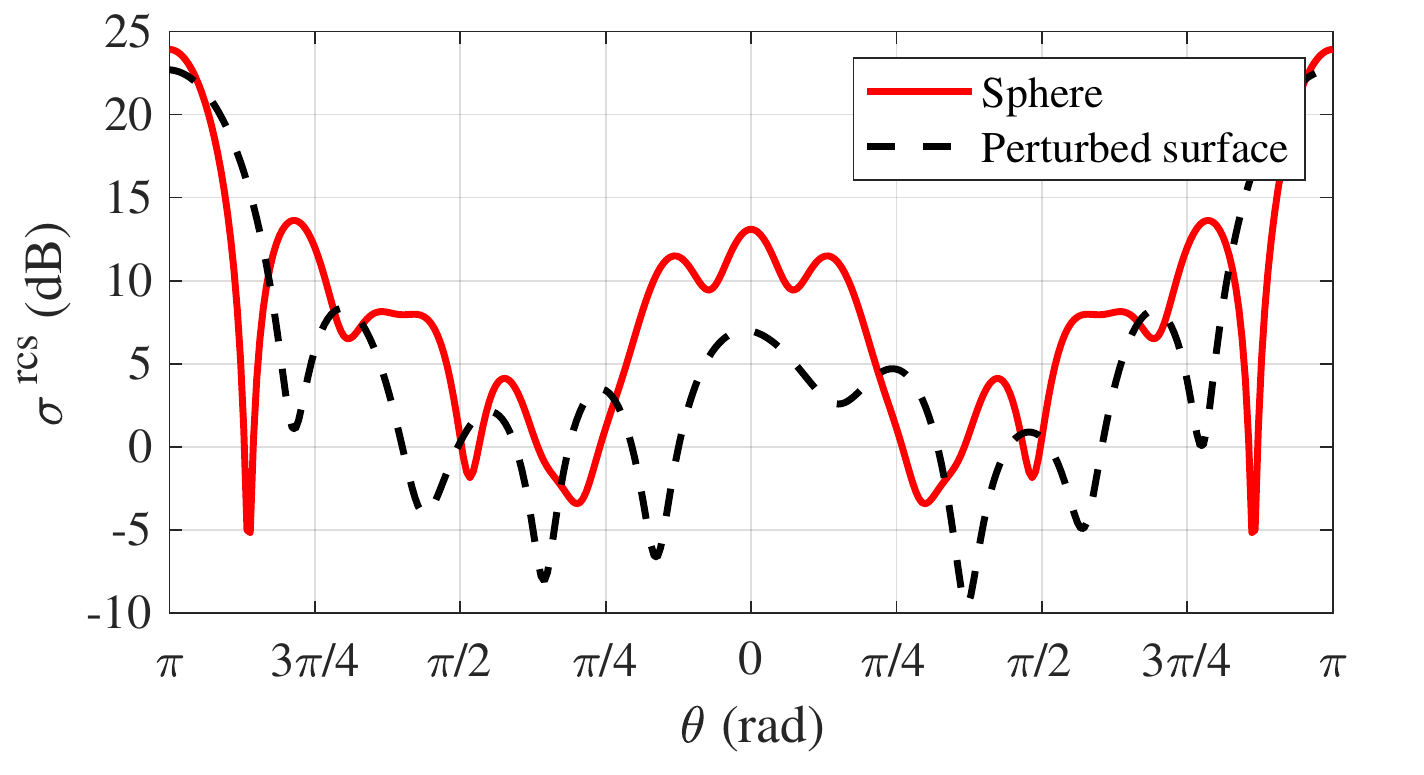}
\label{fig:rcs_yz}
}
\caption{RCS of the unit sphere and the perturbed shape shown in Fig.~\ref{fig:mesh_C} computed on (a) $xz$ and (b) $yz$ planes under excitation by an $\hat {\bf x}$-polarized plane wave propagating in $-\hat {\bf z}$ direction at $300$~MHz. For (a) $\myphi=0$ and $\myphi=\pi$~rad in the first and second halves of the horizontal axis, respectively. For (b), $\myphi=\pi/2$~rad and $\myphi=3\pi/2$~rad in the first and second halves of the horizontal axis, respectively.}
\label{fig:rcs}
\end{figure}

\section{Numerical results}
\label{sec:num_res}
In all numerical experiments considered in this section, the scatterer resides in free space (vacuum) with $\mu_0=4\pi \times 10^{-7}$~H/m and $\eps_0=c_0^2/\mu_0$~F/m, where $c_0$ is the speed of light in vacuum, and the scatterer's permittivity and permeability are $\mu_1=\mu_0$ and $\eps_1=4\eps_0$, respectively. For the plane wave excitation, ${\bf E}_0={\bf \hat x}$, ${\bf \hat u}^{\rm inc}=-\hat {\bf z}$, and $\omega=2\pi f$ with $f=300$~MHz. The QoI is the scattering cross-section $C^{\rm sca}(\Omega)$ computed over the cone $\Omega =[\mytheta_0, \mytheta_1]\times [\myphi_0, \myphi_1]=[1/6, 11/36]\pi \, {\rm rad} \times [5/12, 19/36]\pi \, {\rm rad}$. For all random perturbations, the parameters of the quasi-spherical harmonics are $\co_1=2$, $\co_2=3$, and $\co_3=2$. Meshes $P_{\ell}$ at levels $\ell=0,1,2,3,4$, which are generated using the method described in Section~\ref{sec:RandomPerturb}, have $\{320, 1280, 5120, 20480, 81920\}$ triangles, respectively. Note that not all five mesh levels are required in every experiment. Since the \CMLMC algorithm is stochastic, 15 independent \CMLMC runs are executed for each experiment to statistically characterize the performance of the method.

The matrix system in~\eqref{eq:mom} is solved iteratively using the transpose-free quasi-minimal residual (TFQMR) method~\cite{freund1993} with a tolerance (residual) of $\epsilon_{\ell}=\epsilon_0\beta^{-2\ell}$, where $\ell=0,1,2,3,4$ represent mesh levels, $\epsilon_0=6.0\times10^{-4}$, and $\beta=2$ (Section~\ref{sec:RandomPerturb}). Using a level-dependent tolerance is consistent with the dependence of the discretization error in the QoI, which scales at a (heuristically determined) rate $\Order{h_{\ell}^2} \propto \Order{\beta^{-2\ell}}$ and ensures that the number of iterations is kept in check when analyzing coarser meshes. The integrals in~\eqref{eq:v_m} and~\eqref{eq:z_mn} are computed using the Gaussian quadrature rules; the accuracy of the rule, i.e. number of quadrature points, adjusts with the mesh level: it is seven 
for $\ell=0,1,2$ and six for $\ell=3,4$. For all levels of meshes, the parameters of the FMM-FFT scheme are selected carefully to ensure that it has six digits of accuracy~\cite{chew1995, chew1997}.
All simulations are executed on Intel(R) Xeon(R) CPU E5-2680 v2 $@$ 2.80~GHz DELL workstations with 40 cores and 128~GB RAM. The electromagnetic simulator is implemented in Fortran 90 while the \CMLMC is written in C++ and Python~\cite{cmlmc_gitlib}. The interface between the electromagnetic solver and MLMC library is implemented in C.

\subsection{Single Realization of Random Variables}

This section demonstrates that the scattered fields strongly depend on the shape of the object for the scenarios considered in the numerical experiments, i.e., $f=300$ MHz, $\varepsilon_1=4\varepsilon_0$, and the size of the object is roughly $2$~m. More specifically,  ${\bf J}({\bf r})$ and ${\bf M}({\bf r})$ induced on a (rotated and scaled) perturbed geometry and its RCS are compared to the same quantities of the unit sphere. The rotated, scaled, and perturbed surface is generated using $a_1=0.04$~m, $a_2=0.048$~m, $\myalpha=0.32$~rad, $\mybeta=0.88$~rad, $\mygamma = 0.81$~rad, $l_x=1.06$, $l_y=1.08$, and $l_z=1.07$. The discretization is refined twice, resulting in the mesh with $5120$ triangles [Fig.~\ref{fig:mesh_C}]. A mesh with the same number of triangles is generated on the unit sphere. Figs~\ref{fig:elec_curr} and \ref{fig:mag_curr} show normalized amplitudes of ${\bf J}({\bf r})$ and ${\bf M}({\bf r})$ induced on this unit sphere. Fig.~\ref{fig:elec_curr_pert} and~\ref{fig:mag_curr_pert} plot the normalized amplitudes of ${\bf J}({\bf r})$ and ${\bf M}({\bf r})$ on the perturbed shape. It is clear that there is a significant difference between the current distributions induced on the sphere and the perturbed shaped. Figs.~\ref{fig:rcs_xz} and \ref{fig:rcs_yz} compare $\sigma^{\rm rcs}(\mytheta, \myphi)$ of the unit sphere and the perturbed shaped computed on the $xz-$ ($\mytheta \in [0,\pi]$~rad, $\myphi=0$ and $\myphi=\pi$~rad) and $yz-$ ($\mytheta \in [0, \pi]$~rad, $\myphi=\pi/2$~rad, and $\myphi=3\pi/2$~rad) planes, respectively. 

As expected, the RCS of the perturbed geometry is significantly different than that of the unit sphere. The results presented in Figs.~\ref{fig:currents} and~\ref{fig:rcs} clearly demonstrate the need for a computational tool that can predict EM fields scattered from objects with uncertain shapes parameterized using random variables. 

Below we show two examples with different perturbations of the initial shape. Each of them is illustrated by multiple graphics showing the total work, the average time, the weak and strong convergences, involved levels, and the parameter $\theta$. Additionally, performances of the $\CMLMC$ and MC methods are compared with theoretical estimates. Since the number of mesh points, $N$, is changing from level to level, the computational time and computational work, is shown in terms of $\TOL$, rather than in $N$ (e.g., $\mathcal{O}(\TOL^{-2})$).

\subsection{High variability}\label{ssec:Ex1}
In this example, the \CMLMC algorithm is executed for random variables $a_1$, $a_2\sim \U[-0.14, 0.14]$ m, $\myalpha$, $\mybeta$, $\mygamma \sim \U[0.2,3]$~rad, and $l_x$, $l_y$, $l_z \sim \U[0.9,1.1]$; here $U[a,b]$ is the uniform distribution between $a$ and $b$. The \CMLMC algorithm is run for $\TOL$ values ranging from $0.2$ to $0.008$. At the lowest value of $\TOL$, the \CMLMC algorithm requires five mesh levels, i.e., $P_{\ell}$, $\ell=0,1,2,3,4$.
\begin{figure}[htbp!]
\centering
\subfigure[]{
\includegraphics[page=6, width=0.45\textwidth]{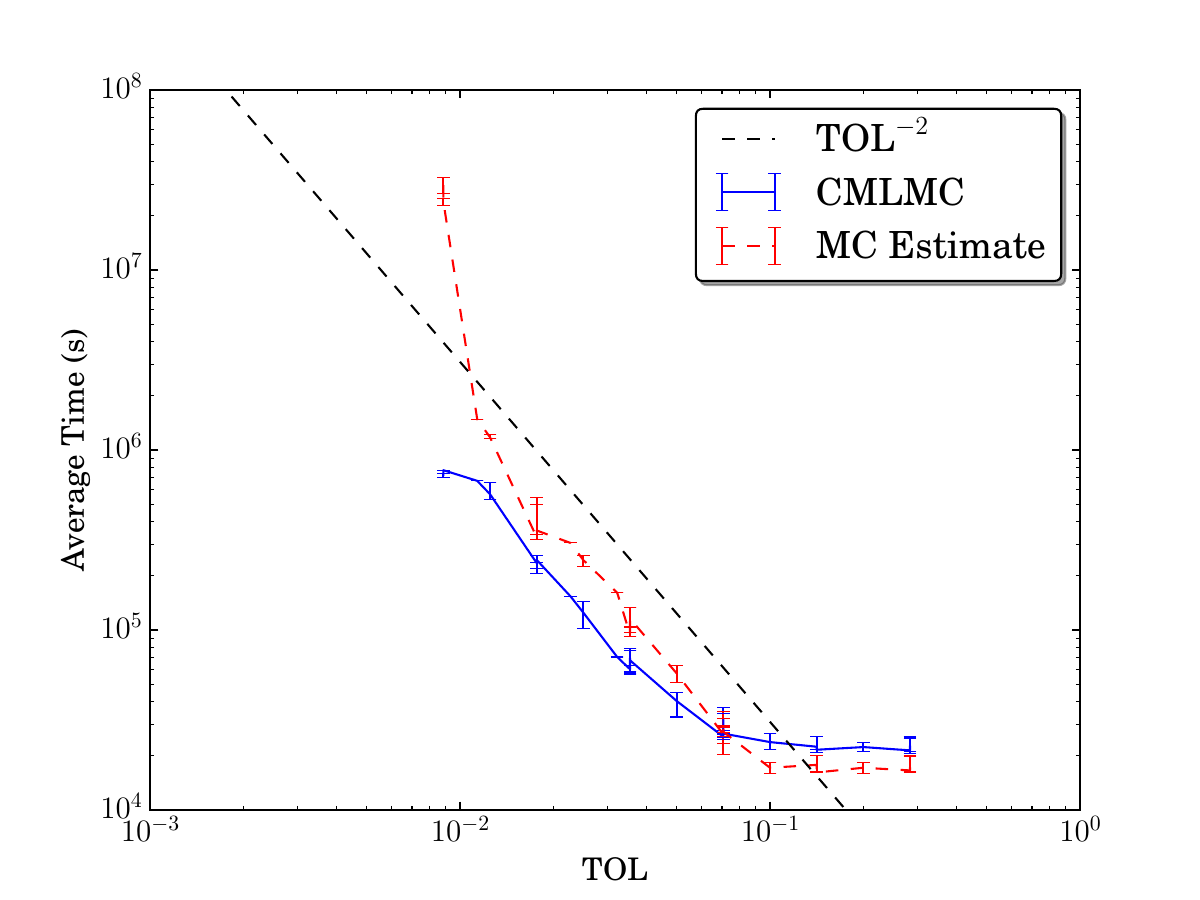}
\label{fig:MLMCpics1-2_A}
}
\subfigure[]{
\includegraphics[page=7, width=0.45\textwidth]{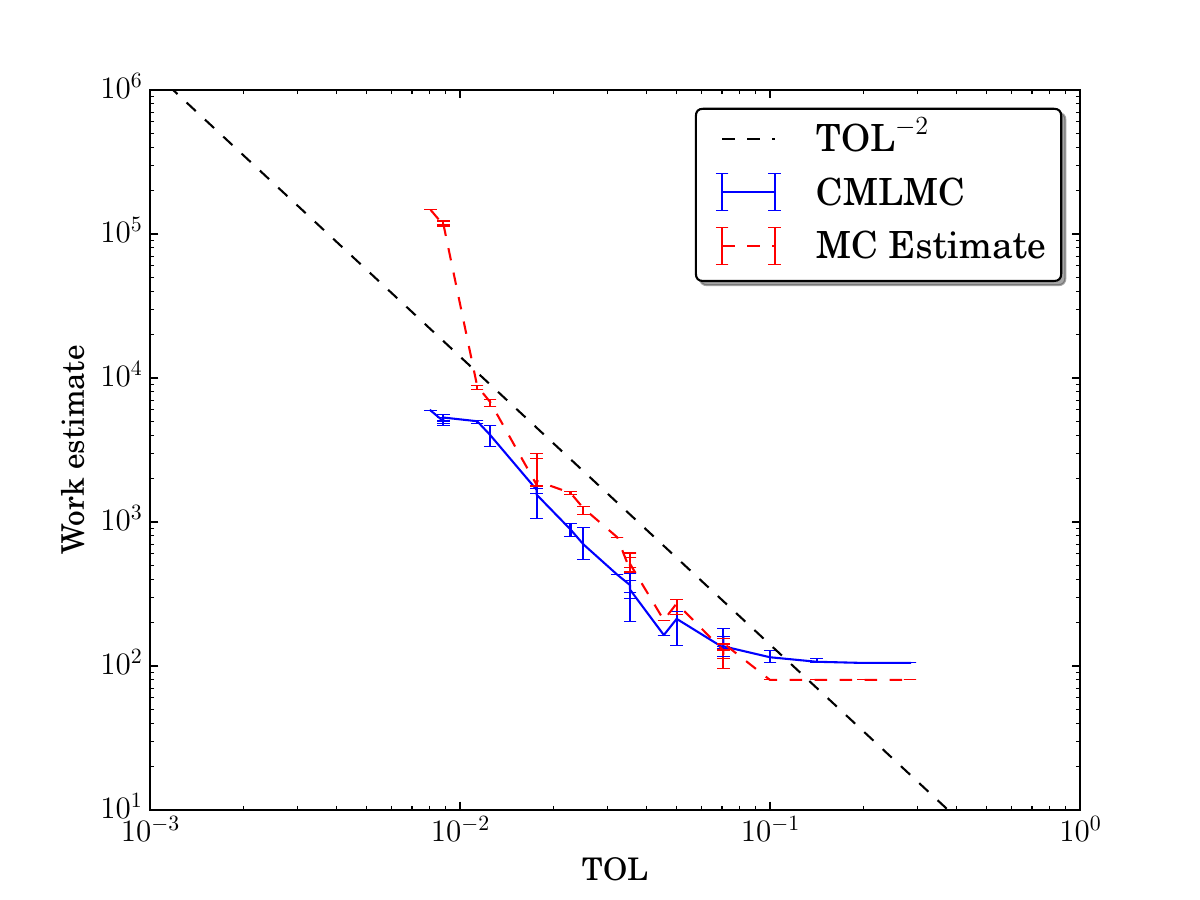}
\label{fig:MLMCpics1-2_B}
}
\caption{(a) Computation times required by the \CMLMC and MC methods vs. $\TOL$. (b) Value of the computational cost estimate $W$ [given by~\eqref{eq:totalwork}] for the \CMLMC and MC methods vs. $\TOL$. Computation times are averaged over 15 repetitions of the experiment.}
\label{fig:MLMCpics1-2}
\end{figure}

Fig.~\ref{fig:MLMCpics1-2_A} compares computation times for the \CMLMC and MC algorithms as a function of $\TOL$. The figure reveals three convergence regimes ($\TOL$ zones).

The first zone, $ 0.1\lessapprox \TOL \lessapprox 0.3$, covers the range where $\TOL$ is larger than the sum of bias and statistical error. No additional samples and no additional mesh refinements are required by the \CMLMC algorithm. The MC and \CMLMC methods consume similar computational resources.

The second zone, $ 0.012\lessapprox \TOL \lessapprox 0.1$, corresponds to a pre-asymptotic regime. Both convergence rates are approximately 2. The statistical error is dominant, and many new samples on existing meshes $P_\ell$, $\ell=0,1,2$ (by default) are required. The MC and \CMLMC methods again consume similar computational resources. To achieve higher accuracy, both the bias and the statistical error should be reduced. The statistical error could be reduced by taking more samples, and the bias by using finer meshes (i.e., increasing $\ell$).

The third zone, $ 0.008\lessapprox \TOL \lessapprox 0.012$, corresponds to the start of the asymptotic regime. The bias becomes important and finer meshes are required. MC computation time increases rapidly and for the smallest $\TOL$ used, the \CMLMC algorithm becomes
almost 10 times faster than the MC method. Note that in the first and second regimes the MC method may outperform the $\CMLMC$ algorithm since the latter carries some initial overhead. 

For example, for $\TOL \gtrapprox 0.012$, only $P_\ell$, $\ell=0,1,2,3$ are required, but for $\TOL \lessapprox 0.012$ an additional finer mesh, $P_{\ell}$, $\ell=4$, is required. Sampling on level $\ell=4$ is more expensive; therefore, the MC computation time increases rapidly for $\TOL \lessapprox 0.012$. On the other than, the \CMLMC algorithm requires only very few samples on level $\ell=4$, and is, therefore, significantly faster. 
\begin{figure}[!htbp] 
\centering
\subfigure[]{
\includegraphics[page=8, width=0.45\textwidth]{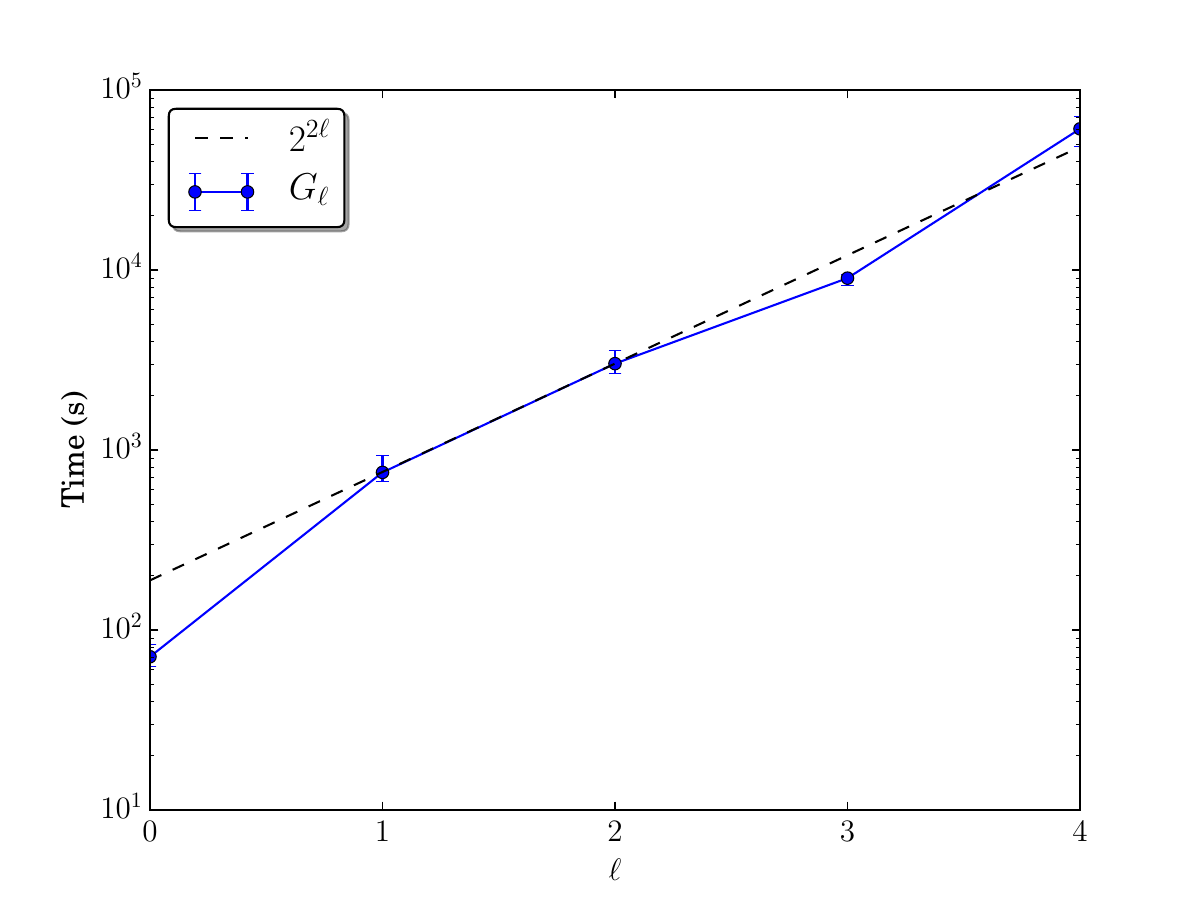}
\label{fig:MLMCpics8-14_A}}
\subfigure[]{
\includegraphics[page=9, width=0.45\textwidth]{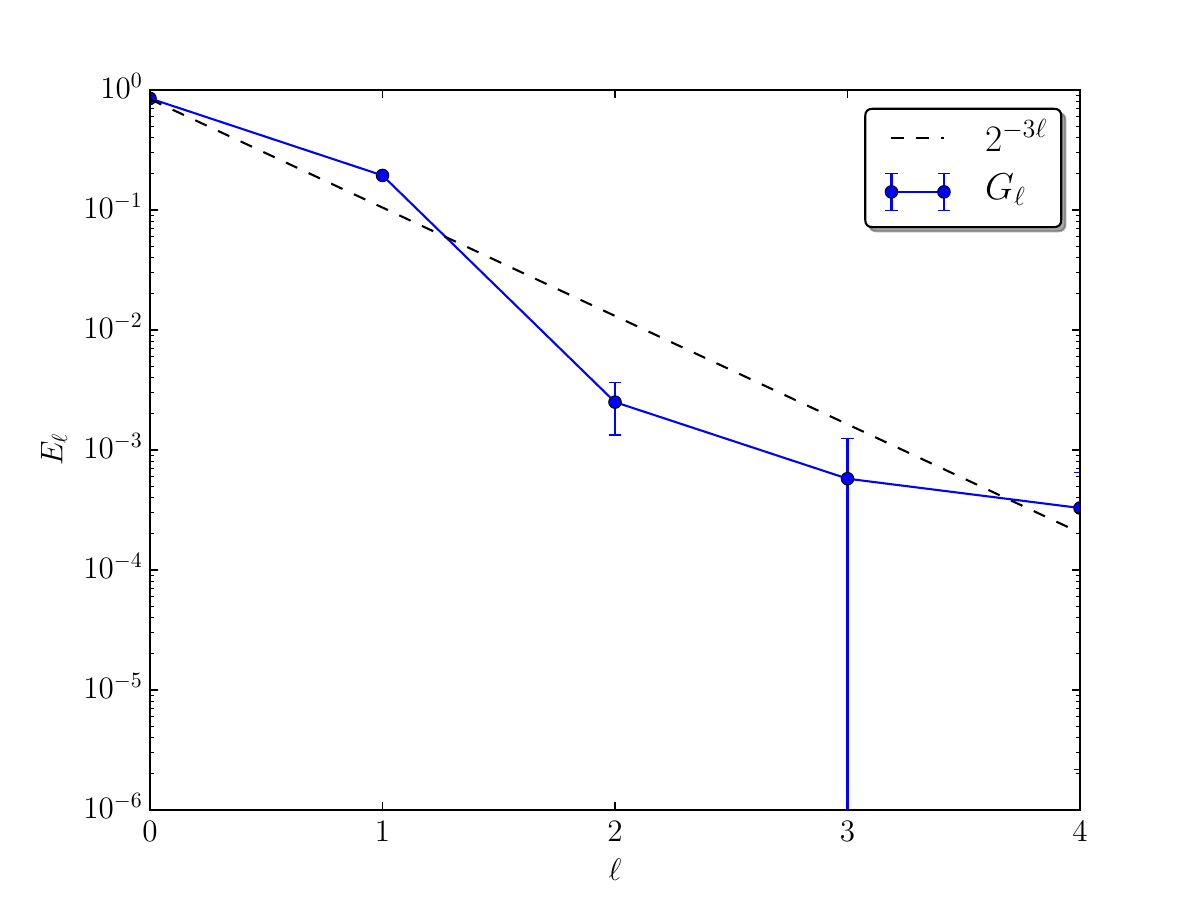}
\label{fig:MLMCpics9-10_A}}\\
\subfigure[]{
\includegraphics[page=10, width=0.45\textwidth]{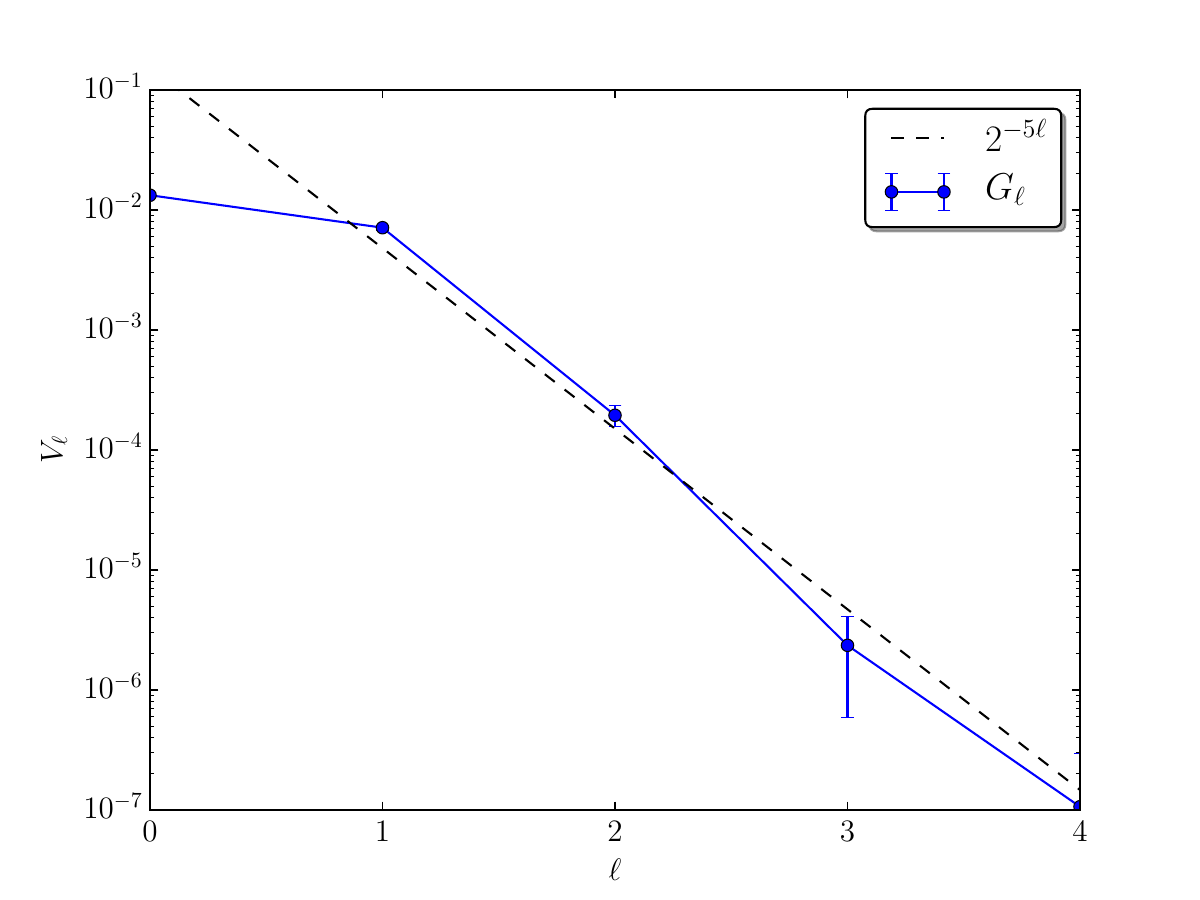}
\label{fig:MLMCpics9-10_B}}
\caption{(a) Time required to compute $G_{\ell}$ vs. ${\ell}$. (b) $E_{\ell}=\E{G_{\ell}}$ vs. $\ell$ and assumed weak convergence curve $2^{-3\ell}$ ($q_1=3$). (c) $V_{\ell}=\var{G_{\ell}}$ vs. $\ell$ and assumed strong convergence curve $2^{-5\ell}$ ($q_2=5$). The experiment is repeated 15 times independently and the obtained values are shown as error bars on the curves.}
\label{fig:MLMCpics9-10}
\end{figure}

The computational cost estimate $W$ is an indicator of computation time. It depends on how the computational cost of the deterministic solver changes from level $\ell-1$ to $\ell$ [as indicated by parameters $\gamma$ and $\beta$ in~\eqref{eq:workpl}] and 
on the order of decay for the mean and the variance [parameters $q_1$, $q_2$ in \eqref{eq:weak_error_model}]. Fig.~\ref{fig:MLMCpics1-2_B} plots the values of $W$ vs. $\TOL$. The curve is similar to that of the \CMLMC computation time given in Fig.~\ref{fig:MLMCpics1-2_A} demonstrating that $\gamma \approx 1$, $q_1 \approx 3$, and $q_2 \approx 5$ are reasonably accurate. 

Fig.~\ref{fig:MLMCpics8-14_A} shows the time required to compute $G_{\ell}=g_{\ell}-g_{\ell-1}$ vs. $\ell$. 
Computation times vary roughly as $2^{2\ell}$, which verifies that $\gamma \approx 1$ [since $d=2$ and $\beta=2$ in~\eqref{eq:workpl}]. Fig.~\ref{fig:MLMCpics9-10_A} shows $E_{\ell}=\E{G_{\ell}}$ vs. $\ell$, revealing that $E_{\ell} \sim 2^{-3\ell}$ (assumed weak convergence obtained with $q_1=3$). Fig.~\ref{fig:MLMCpics9-10_B} shows $V_{\ell}=\var{G_{\ell}}$ vs $\ell$, demonstrating that $V_{\ell}\sim 2^{-5\ell}$ (assumed strong convergence with $q_2=5$). 

The results presented in Figs.~\ref{fig:MLMCpics1-2_A} and~\ref{fig:MLMCpics1-2_B}, and Figs.~\ref{fig:MLMCpics8-14_A}, ~\ref{fig:MLMCpics9-10_A}, and~\ref{fig:MLMCpics9-10_B} confirm the assumptions stated in Section~\ref{sec:MLMC} 
as well as the \CMLMC scheme's quasi-optimality. 
\begin{figure}[!htbp]
\center
\includegraphics[page=14, width=0.45\textwidth]{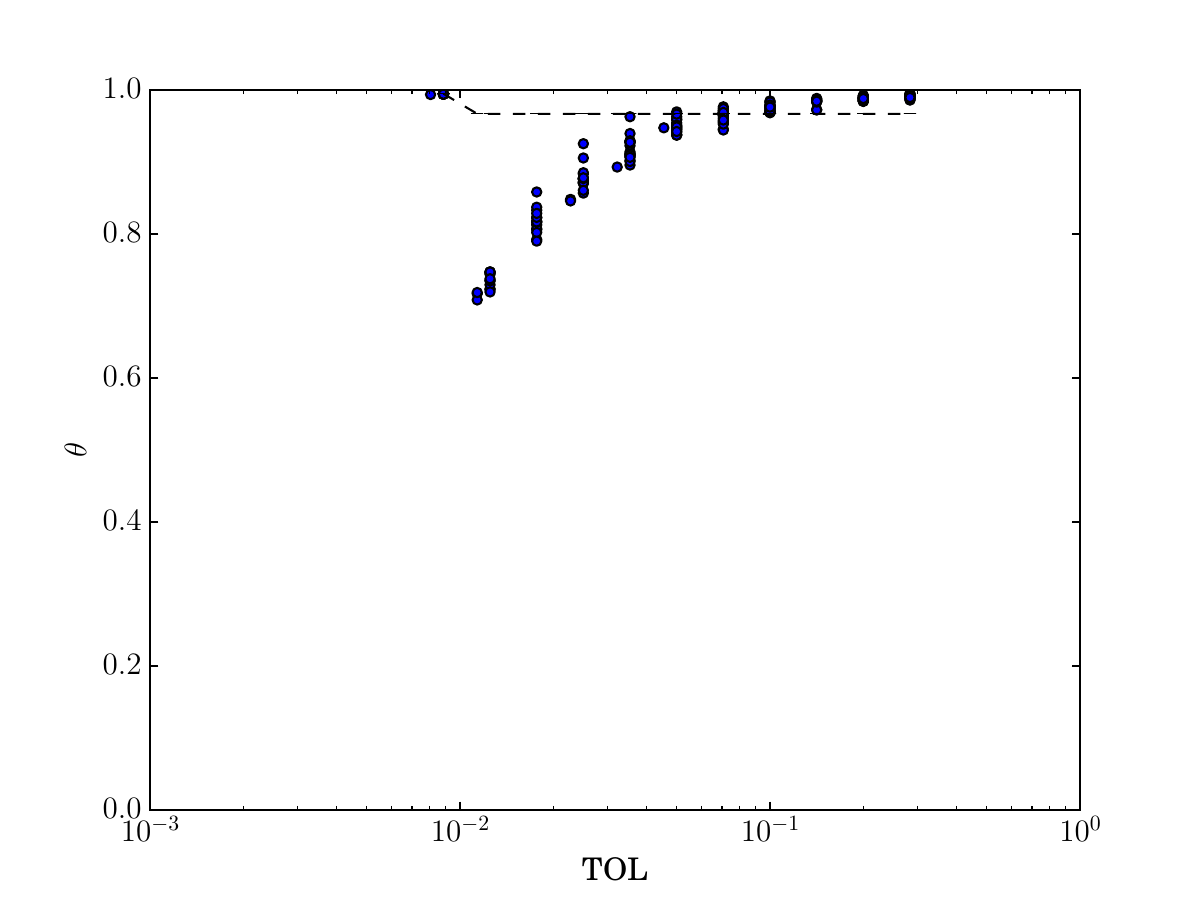}
\caption{Value of $\theta$ used by the \CMLMC algorithm vs. $\TOL$. Computation time is averaged over 15 repetitions of the experiment, i.e., there are 15 values of $\theta$ for a given value of $\TOL$.}
\label{fig:MLMCpics8-14_B}
\end{figure}
\newpage
Fig.~\ref{fig:MLMCpics8-14_B} shows $\theta$ vs. $\TOL$ and demonstrates the complex relationship between $\TOL$ and the bias and statistical error. $\theta$ decreases from 1 to $\approx 0.7$. $\theta=1$ implies that the ratio of bias to the total error is negligible, i.e. that the ratio of of statistical error to the total error is 1. Such variability in $\theta$ is one of the differences between the \CMLMC algorithm and the MLMC method where $\theta=0.5$. The blue dots show that for $\TOL\approx 0.2$, $\theta \approx 1$, meaning that the impact of the bias is negligible and that there is no need to further extend the mesh hierarchy by adding finer mesh levels.

A higher accuracy can be achieved by decreasing either the bias or the statistical error (i.e., smaller $\TOL$). $\theta\approx 1$ means that the bias is negligible and that the statistical error should be decreased to achieve a smaller $\TOL$. Only when introducing a sufficient number of new samples will the statistical error decrease and the ratio of bias to $\TOL$ becomes higher. The bias can be decreased by including an additional mesh level. After that, the ratio of bias to $\TOL$ is dropping again, and the statistical error becomes dominant and should be decreased, etc. For $\TOL \approx 0.01$, the present mesh hierarchy is not sufficient anymore, and the \CMLMC algorithm adds one more mesh level.

\begin{figure}[!htbp]
\centering
\subfigure[]{
\includegraphics[width=0.45\textwidth]{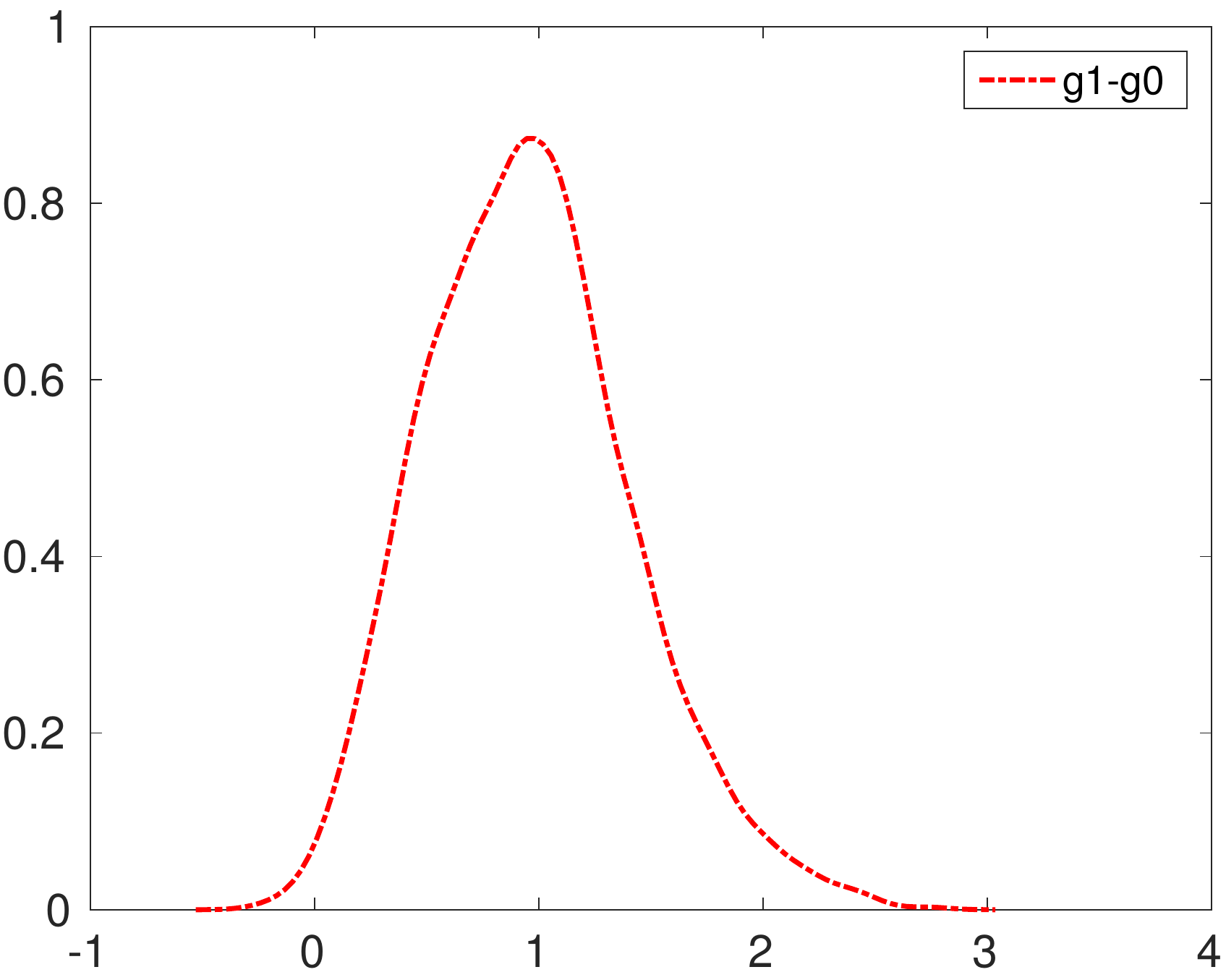}
\label{fig:DensityGl0_A}}
\subfigure[]{
\includegraphics[width=0.45\textwidth]{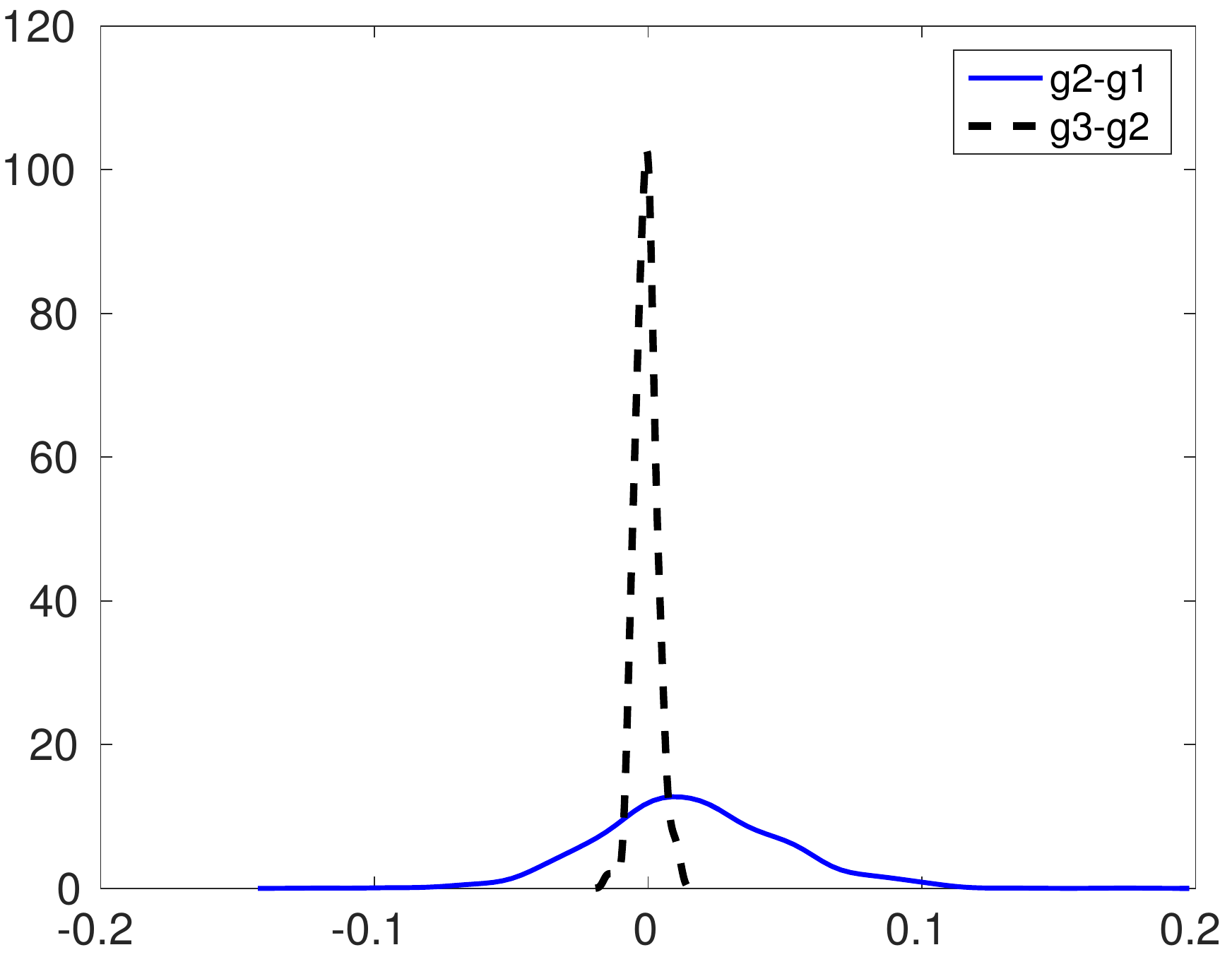}
\label{fig:DensityGl0_B}}
\caption{Pdfs of $g_{\ell}-g_{\ell-1}$ for (a) $\ell=1$ and (b) $\ell=\{2,3\}$. } 
\label{fig:DensityGl0}
\end{figure}

Another way to study the behavior of $G_{\ell}$ is to look at its probability density function (pdf). Figs.~\ref{fig:DensityGl0_A} and~\ref{fig:DensityGl0_B} plot empirical pdfs of $G_{\ell}$ from $\{2000,400,50\}$ samples for $\ell=1$ and $\ell=\{2,3\}$, respectively. Fig.~\ref{fig:DensityGl0_A} shows that $\var{G_{1}}=\var{g_{1}-g_{0}}$, where $g_{0}$ and $g_{1}$ are computed using the meshes $P_0$ and $P_1$, varies roughly in the range $(0, 2.5)$, i.e., this variance is large. Note that $\E{g_{1}-g_{0}}\approx 1.0$. Fig.~\ref{fig:DensityGl0_B} shows that $G_2=g_2-g_1$, where $g_2$ are $g_{1}$ are computed using the meshes $P_2$ and $P_1$, varies in the interval $(-0.1,0.1)$ and $\E{g_2-g_1}\approx 0.02$. Finally, $G_3=g_3-g_2$, where $g_3$ are $g_{2}$ are computed using the meshes $P_3$ and $P_2$, varies in the interval $(-0.02,0.02)$ and $\E{G_3}\approx 0$. The pdfs of $G_{\ell}$ concentrate more and more around zero [with the rates shown in Figs.~\ref{fig:MLMCpics9-10_A}, and~\ref{fig:MLMCpics9-10_B}] as $\ell$ increases. 

\subsection{Low Variability}\label{ssec:Ex2}
In this example, the \CMLMC algorithm is executed for random variables $a_1$, $a_2\sim \U[0.014, 0.49]$ m, $\myalpha$, $\mybeta$, $\mygamma \sim \U[0.2,1.0]$~rad, and $l_x$, $l_y$, $l_z \sim \U[0.8,1.2]$. The variability of these random variables is lower than that of the variables used in the previous example. The value of $\TOL$ is changed from $0.2$ to $0.02$. For the lowest $\TOL$ value, the \CMLMC algorithm requires four mesh levels, i.e., $P_{\ell}$, $\ell=0,1,2,3$ are the only meshes levels used for this experiment.

\begin{figure}[htbp!]
\centering
\subfigure[]{
\includegraphics[page=6,width=0.45\textwidth]{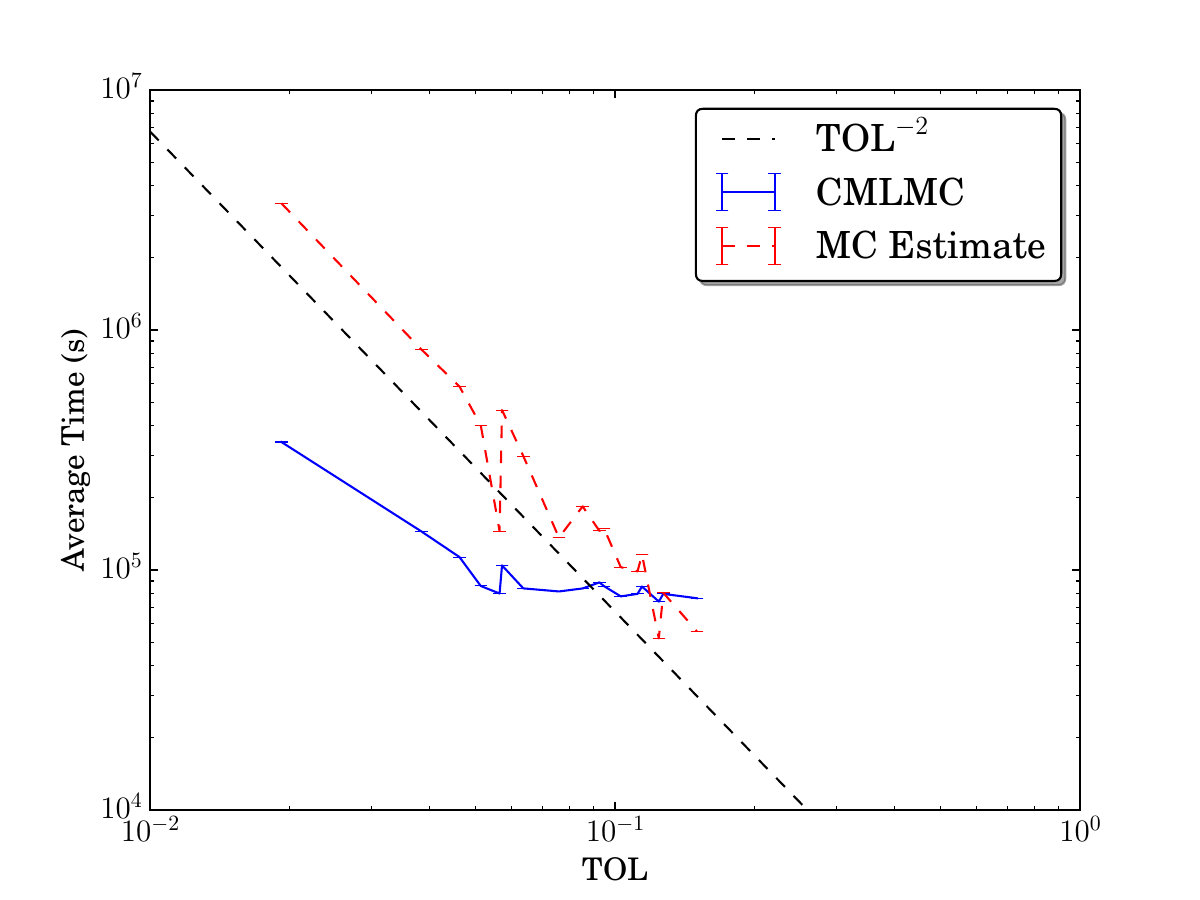}
\label{fig:new3NovA_1-2_A}}
\subfigure[]{
\includegraphics[page=7, width=0.45\textwidth]{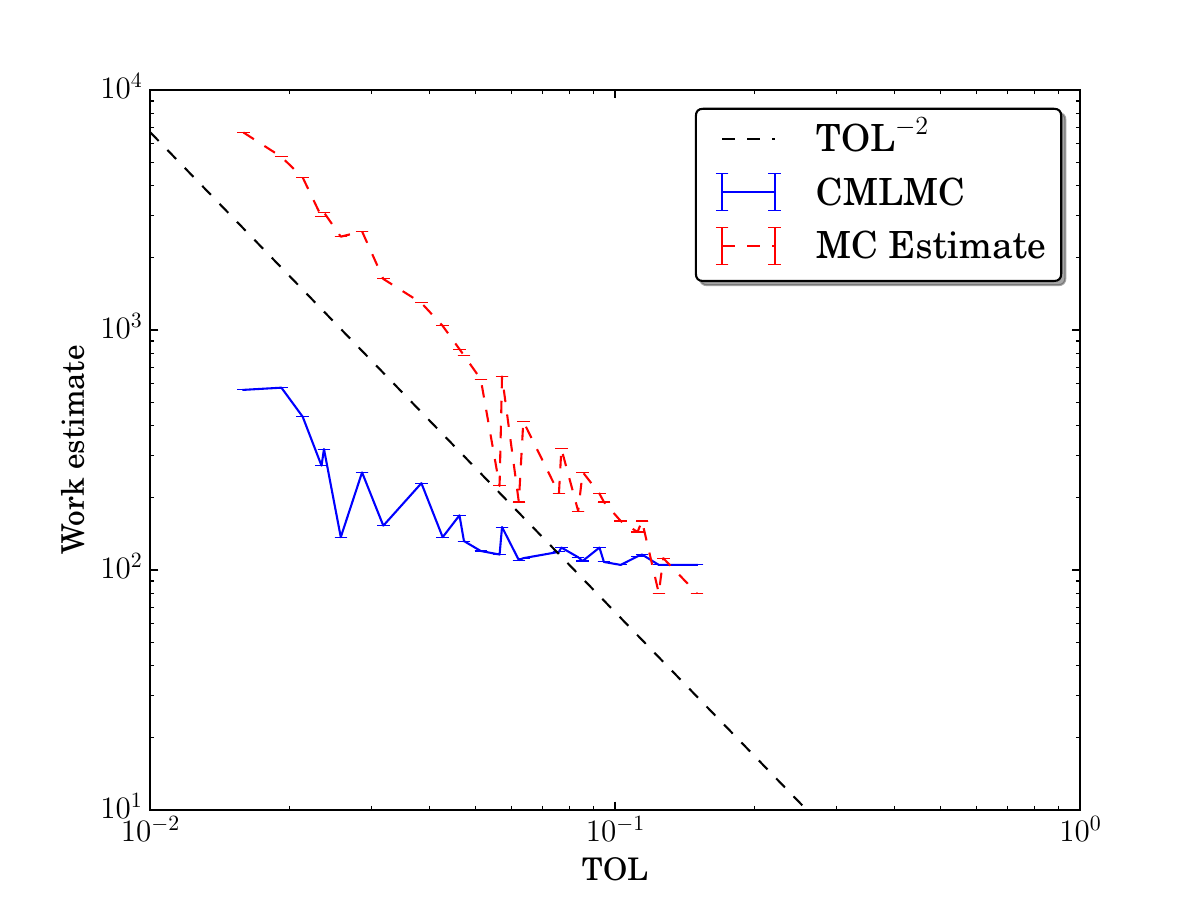}
\label{fig:new3NovA_1-2_B}}
\caption{(a) Computation times required by the \CMLMC and MC methods vs. $\TOL$. (b) Value of the computational cost estimate $W$ [given by~\eqref{eq:workpl}] for the \CMLMC and MC methods vs. $\TOL$. Computations times are averaged over 15 repetitions of the experiment. }
\label{fig:new3NovA_1-2}
\end{figure}

Fig.~\ref{fig:new3NovA_1-2_A} compares the computation time of the \CMLMC and MC methods as a function of $\TOL$.
There are two zones in the figure. The first zone, $ 0.07  \lessapprox \TOL  \lessapprox  0.15$, describes the regime when $\TOL$ is higher than the sum of the bias and statistical error. No additional samples or refinements are needed.  

The second zone, $ \TOL \lessapprox 0.07$, describes the pre-asymptotic regime. As $\TOL$ gets smaller, the \CMLMC algorithm becomes more efficient than the MC method. For values of $\TOL$ close to $0.02$, the \CMLMC algorithm is roughly $10$ times faster than MC. Fig.~\ref{fig:new3NovA_1-2_B} shows the values of the computational cost estimate $W$  vs. $\TOL$. The curve is similar to that of the \CMLMC computation time given in Fig.~\ref{fig:new3NovA_1-2_A} validating the model developed for the \CMLMC algorithm.

\begin{figure}[htbp!]
\centering
\subfigure[]{
\includegraphics[page=9, width=0.45\textwidth]{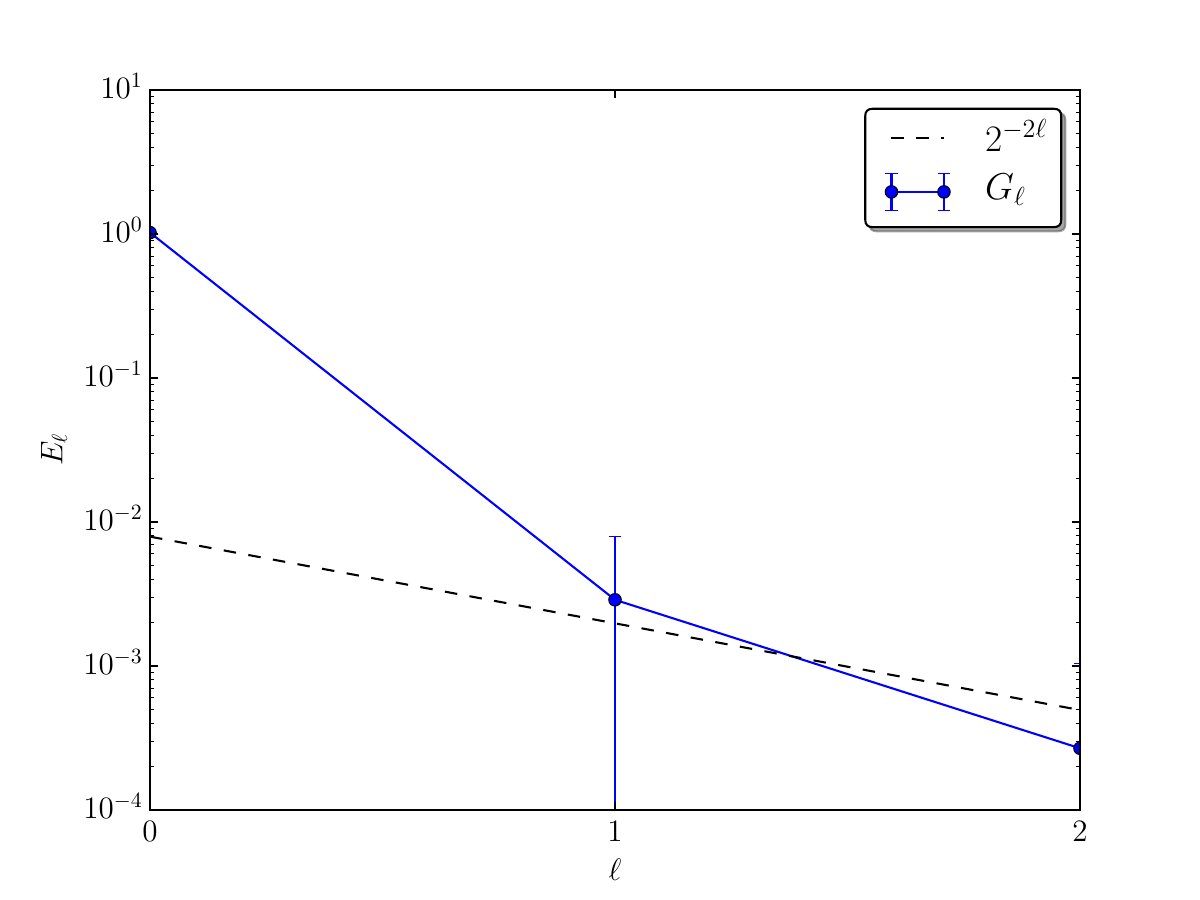}
\label{fig:new3NovA_3-4_A}}
\subfigure[]{
\includegraphics[page=10, width=0.45\textwidth]{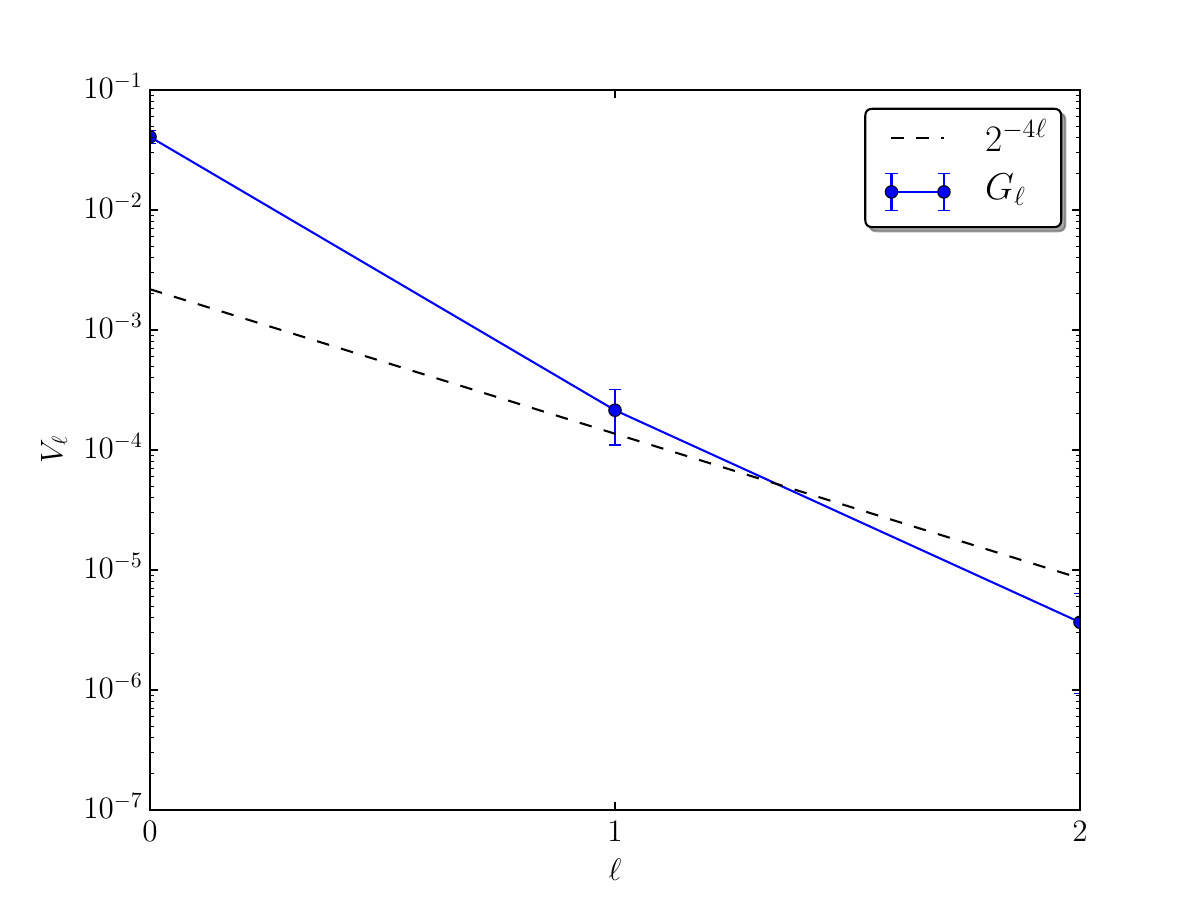}
\label{fig:new3NovA_3-4_B}}
\caption{(a) $E_{\ell}=\E{G_{\ell}}$ vs. $\ell$ and assumed weak convergence curve $2^{-2\ell}$ ($q_1=2$). (b) $V_{\ell}=\var{G_{\ell}}$ vs. $\ell$ and assumed strong convergence curve $2^{-4\ell}$ ($q_2=4$). The experiment is repeated 15 times and error bars are shown on the curves.}
\label{fig:new3NovA_3-4}
\end{figure}

Figs.~\ref{fig:new3NovA_3-4_A} and \ref{fig:new3NovA_3-4_B} show $E_{\ell}=\E{G_{\ell}}$ and  $V_{\ell}=\var{G_{\ell}}$ vs. $\ell$, respectively, revealing dependencies on $\ell$ that vary as $2^{-2\ell}$ (assuming weak convergence obtained with $q_1=2$) and $2^{-4\ell}$ (assuming strong convergence obtained with $q_2=4$). Note that in the previous example these parameters are $3$ and $5$, respectively, demonstrating that the dependence of $q_1$ and $q_2$ on the variability of the random variables used to described the geometry.  

\newpage
\section{Conclusion}
\label{sec:Conclusion}
A computational framework is developed to efficiently and accurately characterize EM wave scattering from dielectric objects with uncertain shapes. To this end, the framework uses the \CMLMC algorithm, which reduces the computational cost of the traditional MC method by performing most of the simulations with lower accuracy and lower cost (using coarser meshes) and smaller number of simulations with higher accuracy and higher cost (using finer meshes). To increase the efficiency further, each of the simulations is carried out using the FMM-FFT accelerated PMCHWT-SIE solver. Numerical results demonstrate that the \CMLMC algorithm can be 10 times faster than the traditional MC method depending on the amplitude of the perturbations used for representing the uncertainties in the scatterer's shape. This work confirms that the known advantages of the \CMLMC algorithm can be observed when it is applied to EM wave scattering: non-intrusiveness, dimension independence, better convergence rates compared to the classical MC method, and higher immunity to irregularity w.r.t. uncertain parameters, than, for example, sparse grid methods.

For optimal performance (for a given value of accuracy parameter \TOL), the \CMLMC algorithm requires the mean and the variance to have reliable convergence rates (i.e., one should be able estimate $q_1$ and $q_2$ without much difference from one level to next). However, some random perturbations may affect the convergence rates. With difficult-to-predict convergence rates, it is hard for the \CMLMC algorithm to estimate the computational cost $W$, the number of levels $L$, the number of samples on each level $M_{\ell}$, the computation time, and the parameter $\theta$, and the variance in QoI. All these may result in a sub-optimal performance. Indeed, numerical results demonstrate that there is a significant pre-asymptotic regime where the performance is not optimal.  Additionally, it is observed that the settings of the FMM-FFT accelerated PMCHWT-SIE solver, which regulate the computation time and the accuracy (such as the iterative solver threshold, the number of quadrature points, and the FMM-FFT parameters), have a significant effect on the performance of the \CMLMC algorithm. 

It should also be noted here that the \CMLMC algorithm is proven effective not only in studying the effects of uncertainty in the geometry on scattered fields but also in increasing the robustness of the FMM-FFT accelerated PMCHWT-SIE solver by testing its convergence for a large set of scenarios with deformed geometries and varying mesh densities, quadrature rules, iterative solver tolerances, and FMM parameters.

%
%

\section*{Acknowledgements}
The research reported in this publication was supported by funding from King 
Abdullah University of Science and Technology (KAUST). We gratefully
acknowledge support from Abdul-Lateef Haji-Ali (KAUST and Oxford), developer of the \CMLMC, Erik von Schwerin and Haakon Hoel for valuable comments, concerning the \CMLMC algorithm, as well as
Ismail Enes Uysal and H\"useyin Arda \"Ulk\"u (KAUST) for their initial efforts on this project. 


\end{document}